\documentclass[nature, footinbib,floatfix,superscriptaddress]{revtex4-1}

\usepackage{graphics,bm,amsmath,amssymb,natbib,url,epsfig,psfrag,color,tikz,pgf, booktabs,multirow, physics}
\usetikzlibrary{arrows,calc,shapes,decorations.pathreplacing}
\usepackage{mathtools}
\usepackage{babel}
\usetikzlibrary{shapes, calc, arrows} %

\begin{document}

\title{Non-Abelian Anyon Collider}

\author{June-Young M. Lee}  
\affiliation{Department of Physics, Korea Advanced Institute of Science and Technology, Daejeon 34141, Korea}
\author{H.-S. Sim} \email[]{hssim@kaist.ac.kr}
\date{\today}
\affiliation{Department of Physics, Korea Advanced Institute of Science and Technology, Daejeon 34141, Korea}

\begin{abstract}
A collider, where particles are injected to a beam splitter from opposite sides, has been used for identifying quantum statistics of identical particles. 
The collision leads to bunching of the particles for bosons and antibunching for fermions \cite{HOM, Liu98}.
In recent experiments~\cite{Bartolomei20}, a collider was applied to a fractional quantum Hall regime hosting Abelian anyons.
The observed negative cross correlation of electrical currents cannot be understood with fermionic antibunching~\cite{Rosenow16}.
Here we predict, based on a conformal field theory and non-perturbative treatment of non-equilibrium anyon injection, that the collider provides a tool for direct observation of the braiding statistics of various Abelian and non-Abelian anyons. Its dominant process is not direct collision between injected anyons, contrary to common expectation, but braiding between injected anyons and an anyon excited at the collider.
The dependence of the resulting negative  cross correlation on the injection currents distinguishes non-Abelian SU(2)$_k$ anyons, Ising anyons, and Abelian Laughlin anyons.
\end{abstract}

\let\clearpage\relax
\maketitle

Anyons are quasiparticles that are neither fermions nor bosons~\cite{Leinaas, Arovas}.  
They exhibit fractional statistics behavior when an anyon winds around another in two dimension. 
This is characterized by the overlap, called monodromy, between 
their states before and after the winding or braiding~\cite{Bonderson08}. While bosons and fermions have the trivial monodromy $M=1$, 
Abelian anyons have a complex phase factor $M = e^{-i 2 \theta}$, where $\theta \ne 0, \pi$ is their position exchange phase.
Non-Abelian anyons have a monodromy of $|M| < 1$, as their braiding results in unitary rotation of their state in a degenarate state manifold. 
The unitary rotation is an element of topological quantum computing~\cite{Nayak}. 
It is expected that along fractional quantum Hall edge channels there flow anyons
such as Abelian Laughlin anyons at filling factor $\nu = 1/3$, non-Abelian SU(2)$_{k=2}$ anyons of the anti-Pfaffian state \cite{Lee07,Levin07} or Ising anyons of the particle-hole Pfaffian state at $\nu = 5/2$ \cite{Zucker},
and non-Abelian SU(2)$_{k=3}$ anyons of the anti-Read-Rezayi state at $\nu = 12/5$ \cite{Bishara08-ARR}.

\begin{figure}[t]
	\centering
	\includegraphics[width =  1.\textwidth]{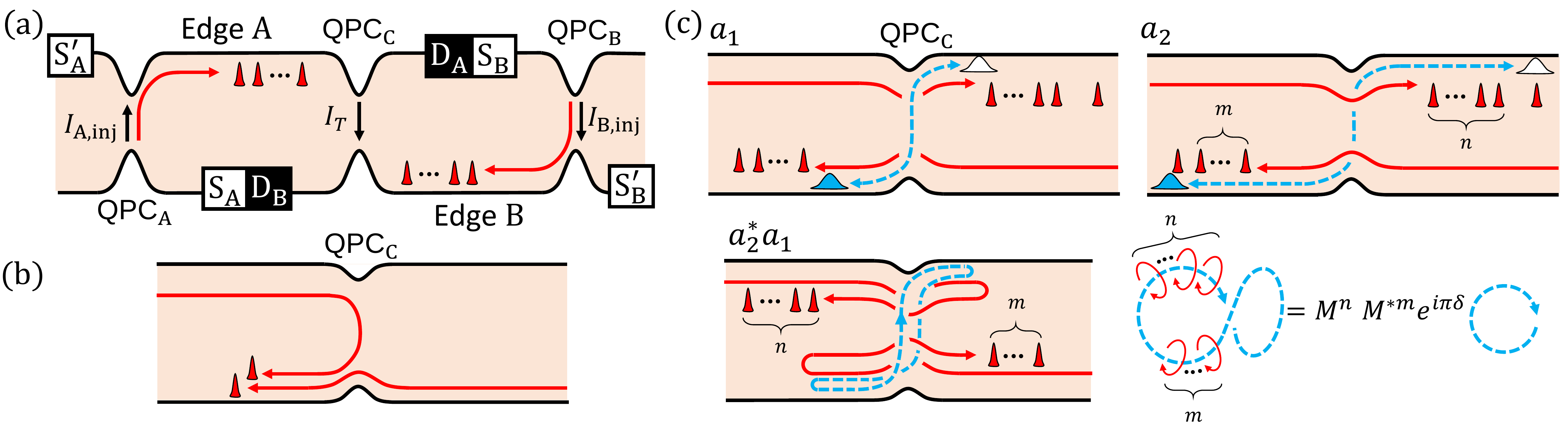}
	\caption{ Fractional quantum Hall collider. (a)  Setup.  
Anyons are injected to Edge A/B through QPC$_\textrm{A/B}$ by voltage $V_\textrm{A/B,inj}$ applied to Source S$_\textrm{A/B}$, accompanied by current $I_\textrm{A/B,inj}$ of charge $e^*$. The injected anyons (red narrow peaks) flow downstream to QPC$_\textrm{C}$ (red trajectories); a corresponding setup for upstream anyons is shown in Supplementary Material. The QPCs are in the weak backscattering regime. (b) Conventional collision where an injected anyon collides with another after tunneling at QPC$_\textrm{C}$. (c) Time-domain interference involving $(n,m)$ braiding. Its subprocesses $a_1$ and $a_2$ share the common %
spatial locations of injected anyons on the Edges. They have tunneling of an additional anyon at QPC$_\textrm{C}$  (blue wide peaks for the anyon, white peaks for its hole counterpart) by thermal or quantum fluctuations, but at different times (blue trajectories).
In $a_1$ (resp. $a_2$), the tunneling happens after (resp. before) $n$ and $m$ injected anyons pass QPC$_\textrm{C}$ on Edges A and B. 
In their interference $a_2^* a_1$, the additional anyon braids the injected anyons, depicted as a blue twisted loop topologically linked with $n$ ``counterclockwise'' and $m$ ``clockwise'' red loops. Untying  and untwisting the loops give monodromy $M^n (M^*)^m$ and topological spin $e^{i \pi \delta}$.
}\label{fig1}
\end{figure}

On top of long time efforts~\cite{de-Picciotto97, Saminadayar97, Dolev08, Chamon97, Bishara08, Stern06,Bonderson06, Willett09, Ofek10, An11, Rosenow16, Han16, Blee19, Fendley07, Banerjee18, Kane03, Rosenow09, Bishara09, Keyserlingk} on detecting the fractional statistics, there were experimental breakthoughs at $\nu = 1/3$~\cite{Bartolomei20,Nakamura20}. 
In a collider experiment~\cite{Bartolomei20},
two dilute streams of Abelian anyons 
are injected to a quantum point contact (QPC) that behaves as a collider beam splitter [see Fig.~\ref{fig1}(a)]. 
It shows negative cross correlations of electrical currents at the output ports of the collider in agreement with a nonequilibrium bosonization theory~\cite{Rosenow16}. 
It however remains unclear which aspect of the Abelian anyon statistics is identified from the experimental result.
On one hand, it seems natural to interpret the result as an intermediate between fermionic antibunching and bosonic bunching by direct collision [Fig.~\ref{fig1}(b)] between injected anyons~\cite{Rosenow16}.
On the other hand, a braiding effect  
was predicted~\cite{Han16,Blee19} in a related setup where Abelian anyons are injected from only one side.
The identification is important in pursuing a more direct evidence of anyons.
It is also intriguing to apply the collider to non-Abelian anyons. 
There has been no prediction on this issue.

We here develop a theory of a collider encompassing generic Abelian and non-Abelian anyons in fractional quantum Hall systems. 
We demonstrate that for Abelian and non-Abelian anyons, its dominant process is ``time-domain'' interference,
in which an anyon, excited at the collider QPC$_\textrm{C}$, braids the injected anyons passing QPC$_\textrm{C}$ within the interference time window. More anyons are braided as more injected anyons pass. So the cross correlation depends on the product of the injection current and the monodromy from the braiding, differentiating various anyons. The interference is absent in bosons and fermions, where it corresponds to a trivial vacuum bubble process that does not contribute to observables.  Hence the dependence cannot be interpreted as deviation from fermionic antibunching due to the commonly anticipated direct collision between injected anyons.

In Fig.~\ref{fig1}(a), anyons are injected with rate $I_\text{A/B,inj}/e^*$ at QPC$_\textrm{A/B}$ by voltage $V_\text{A/B,inj}$, and flow to QPC$_\textrm{C}$.
They are downstream charged anyons or upstream charge-neutral anyons. Anyon tuneling at QPC$_\text{C}$ is described by Hamiltonian $H_\textrm{T} = \mathcal{T}(t) + \mathcal{T}^\dagger(t) = \gamma_\text{C}[\psi^\dagger_\textrm{B}(0,t)\psi_\textrm{A}(0,t)]_I + \text{h.c.}$. $\gamma_\textrm{C}$ is the tunneling strength, $\psi^\dagger_\textrm{A/B}(x,t)$ creates an anyon on Edge A/B at position $x$ and time $t$, and $[\cdots]_I$ indicates the vacuum fusion channel of the anyon.
We consider the dilute injection of $e^* V_\text{A/B,inj} \gg h I_\text{A/B,inj}/e^*$ in non-equilibrium  with $e^* V_\text{A/B,inj} \gg k_B T$ at temperature $T$ as in experiments~\cite{Bartolomei20}, and derive the non-equilibrium correlator of the tunneling operators   
\begin{equation}\label{neqtunneling}
\begin{split}
\expval{\mathcal{T}^\dagger(0)\mathcal{T}(t)}_\text{neq} = & e^{- \mathcal{I} t}	\expval{\mathcal{T}^\dagger(0)\mathcal{T}(t)}_\text{eq} + \textrm{subleading terms}, \\
\mathcal{I} = & (1-M)  \frac{I_\text{A,inj}}{ e^*} + (1-M^*) \frac{I_\text{B,inj}}{ e^*} = \Re[1-M] \frac{I_+}{ e^*} + i\Im[1-M] \frac{I_-}{ e^*} 
\end{split}
\end{equation}
for $t>0$, using the conformal field theory (CFT), Keldysh nonequilibrium theory, 
and non-perturbative resummation over all perturbation orders of anyon tunneling at QPC$_\textrm{A/B}$ (Supplementary Matarial);
for $t <0$,  $t \to -t$ and $M \to M^*$  are replaced in Eq.~\eqref{neqtunneling}.
$\expval{\cdots}_\text{eq}$ is the equilibrium correlator at $V_\text{A/B,inj} = 0$ and $I_{\pm} = I_\text{A,inj} \pm I_\text{B,inj}$.
Equation~\eqref{neqtunneling} is valid at $t \gg \hbar/e^*V_\text{A/B,inj}$.

The current $I_\textrm{T}$ and its zero-frequency noise $\expval{\delta I_\textrm{T}^2}$ at QPC$_\text{C}$ are written as
$I_\textrm{T} = e^* \int_{-\infty}^\infty dt \expval{\comm{\mathcal{T}^\dagger(0)}{\mathcal{T}(t)}}_\text{neq}$ and $\expval{\delta I_\textrm{T}^2} = e^{*2}\int_{-\infty}^\infty dt \expval{\acomm{\mathcal{T}^\dagger(0)}{\mathcal{T}(t)}}_\text{neq}$
in the lowest tunneling order $O(|\gamma_\textrm{C}|^2)$ at QPC$_\text{C}$.
Using Eq.~\eqref{neqtunneling}, we find
\begin{equation} \label{generalI}
\begin{split}
I_\textrm{T}= &	-4e^*\abs{\gamma_\text{C}}^2 d_\psi^{-1} \Gamma(1-2\delta) \sin\pi\delta	\Im\mathcal{I}^{2\delta-1} + \textrm{subleading terms},  \\
\expval{\delta I_\textrm{T}^2} = & 4e^{*2}\abs{\gamma_\text{C}}^2d_\psi^{-1} \Gamma(1-2\delta) \cos \pi \delta \Re \mathcal{I}^{2\delta -1} + \textrm{subleading terms} 
\end{split}
\end{equation}
at $e^* V_\text{A/B,inj} \gg h I_\text{A/B,inj}/e^*$ and zero temperature (see Methods for finite temperature).
$d_\psi$ and $\delta$ are the quantum dimension and tunneling exponent of the anyon, and $\Gamma$ is the gamma function.  %
The zero-frequency cross correlation $\expval{\delta I_\textrm{A} \delta I_\textrm{B}}$ of the collider output currents at Detectors D$_\textrm{A}$ and D$_\textrm{B}$ is related with $I_\textrm{T}$ and $\expval{\delta I_\textrm{T}^2}$ (Methods). 

It is remarkable that the observables depend on the product $\mathcal{I}$ of the injection currents $I_\text{A/B,inj}$ and the  monodromy factor $(M-1)$ in Eq.~\eqref{neqtunneling}.
Its origin, the time-domain interference involving anyon braiding, is identified, using our perturbation approach.
We consider  an interference event $(n,m)$ between two subprocesses $a_1$ and $a_2$ in a time window $t$. 
Tunneling of an anyon happens at QPC$_\textrm{C}$ at time $t$ in $a_1$ and at time $0$ in $a_2$ [Fig.~\ref{fig1}(c)].
This tunneling occurs not by the voltage $V_\text{A/B,inj}$ but by thermal excitations, and it is described by the equilibrium correlator $\expval{\mathcal{T}^\dagger (0) \mathcal{T}(t)}_\text{eq}$ in Eq.~\eqref{neqtunneling}.   Within the time window, $n$ anyons on Edge A and $m$ anyons on Edge B pass QPC$_\textrm{C}$ without tunneling.
These anyons were injected by $V_\text{A/B,inj}$.
So the interference loop $a_2^* a_1$ in the time axis braids the $n$ anyons on Edge A in a direction and the $m$ anyons on Edge B in the opposite direction, gaining monodromy $M^n (M^*)^m$.
The braiding happens with probability $p_\text{A}(n,t) p_\text{B}(m,t)$ where $p_\alpha(n_\alpha,t)= (\bar{n}^{n_\alpha})/n_\alpha!)e^{-\bar{n}}$ is the Poisson probability distribution for random anyon injections $n_\alpha$ times at QPC$_{\alpha = \text{A,B}}$ over time $t$, with average number $\bar{n} (t,\alpha) = I_{\alpha,\text{inj}} t/e^*$.
Average of the monoromy over different $(n,m)$'s reproduces the exponential factor in Eq.~\eqref{neqtunneling},
\begin{equation}\label{nm_order}  
\exp(\frac{I_\text{A,inj}}{e^*}(M-1)t+\frac{I_\text{B,inj}}{e^*}(M^*-1)t) = \sum_{n,m} p_\text{A}(n,t)p_\text{B}(m,t) M^n (M^*)^m.
\end{equation}
The validity condition of Eq.~\eqref{generalI} with large $V_\text{A/B,inj}$
is necessary for the braiding; the temporal width $h/(e^*V_\text{A/B,inj})$ of the injected anyons must be narrower than their separation $e^* / I_\textrm{A/B,inj}$ and the window $t \lesssim h / (k_B T)$.  
The braiding happens even when anyons are injected from only one side, $I_\textrm{A,inj} = 0$ or $I_\textrm{B,inj} = 0$. 
 
The time-domain interference is distinct from the conventional collision in Fig.~\ref{fig1}(b). 
In the former, the anyon tunneling at QPC$_\textrm{C}$ occurs thermally.
In the latter an anyon injected by  the voltage $V_\text{A/B,inj}$ undergoes tunneling at QPC$_\textrm{C}$.
The former dominates over the latter at $e^*V_\text{A/B,inj} \gg k_B T$ and determines Eq.~\eqref{generalI},
when the tunneling exponent $\delta$ of QPC$_\textrm{C}$ is smaller than 1 (Supplementary Materials).
This is implied from the voltage dependence $I \sim V^{2 \delta -1}$ of QPC tunneling currents in the fractional quantm Hall regime.
We note that the factors $\sin \pi \delta$ and $\cos \pi \delta$ in Eq.~\eqref{generalI} come from the topological spin  or twist factor~\cite{Bonderson08}  $e^{i \pi \delta} = e^{i 2 \pi h_\psi}$ that appears due to operator ordering exchange in the equilibrium correlator $\expval{\mathcal{T}^\dagger (0) \mathcal{T}(t)}_\text{eq}$ for the anyon excited at QPC$_\textrm{C}$, where $h_\psi$ $(= \delta /2)$ is the scaling dimension of the anyon. For Abelian anyons, $e^{i\pi\delta }$ coincides with the exchange phase $e^{i\theta}$.

The dependence of the observables on the product $\mathcal{I}$ in Eq.~\eqref{generalI} offers possibilities of directly observing anyon braiding. The Fano factor $P_- (I_-/I_+) \equiv  \frac{\expval{\delta I_\textrm{A} \delta I_\textrm{B}}}{e^*I_+ \frac{\partial I_\textrm{T}}{\partial I_-}|_{I_-=0}}$ introduced in Ref.~\cite{Rosenow16} is useful.
When $I_\text{A,inj} = I_\text{B,inj}$, we find 
\begin{equation}\label{oldfano}
\begin{split}
P_- (0) =  1 - \frac{\textrm{Re} [1-M]}{\textrm{Im} [1-M]} \frac{\cot \pi \delta }{1-2 \delta}
\end{split}  
\end{equation}
at zero temperature.
For Abelian anyons, $M= e^{-2 i \theta}$, then Eq.~\eqref{oldfano} becomes identical to the expression that was found in Ref. \cite{Rosenow16} but without recognition of the braiding.
The dependence of $P_-$ on $I_-/I_+$ was observed at $\nu = 1/3$~\cite{Bartolomei20}. 
Our time-domain interference implies that the observation is an evidence of Abelian anyon braiding.
 
Our findings are equally applicable to non-Abelian anyons.
At $\nu = 5/2$ or $12/5$, tunneling at QPCs generates downstream Abelian anyons and upstream non-Abelian anyons together,
hence, one can inject the former or latter selectively to QPC$_\textrm{C}$, to observe its braiding; side effects by back flows from QPC$_\textrm{C}$ to QPC$_\textrm{A/B}$ are negligible in our paremeter regime (Supplementary Material).
In Eqs.~\eqref{generalI} and \eqref{oldfano}, $M$ is  the monodromy of the injected anyon, while $\delta$ is the tunneling exponent of  the composite of the former and latter, since they together tunnel at QPC$_\textrm{C}$.
  
\begin{figure}[b]
	\centering
	\includegraphics[width = 0.26 \textwidth]{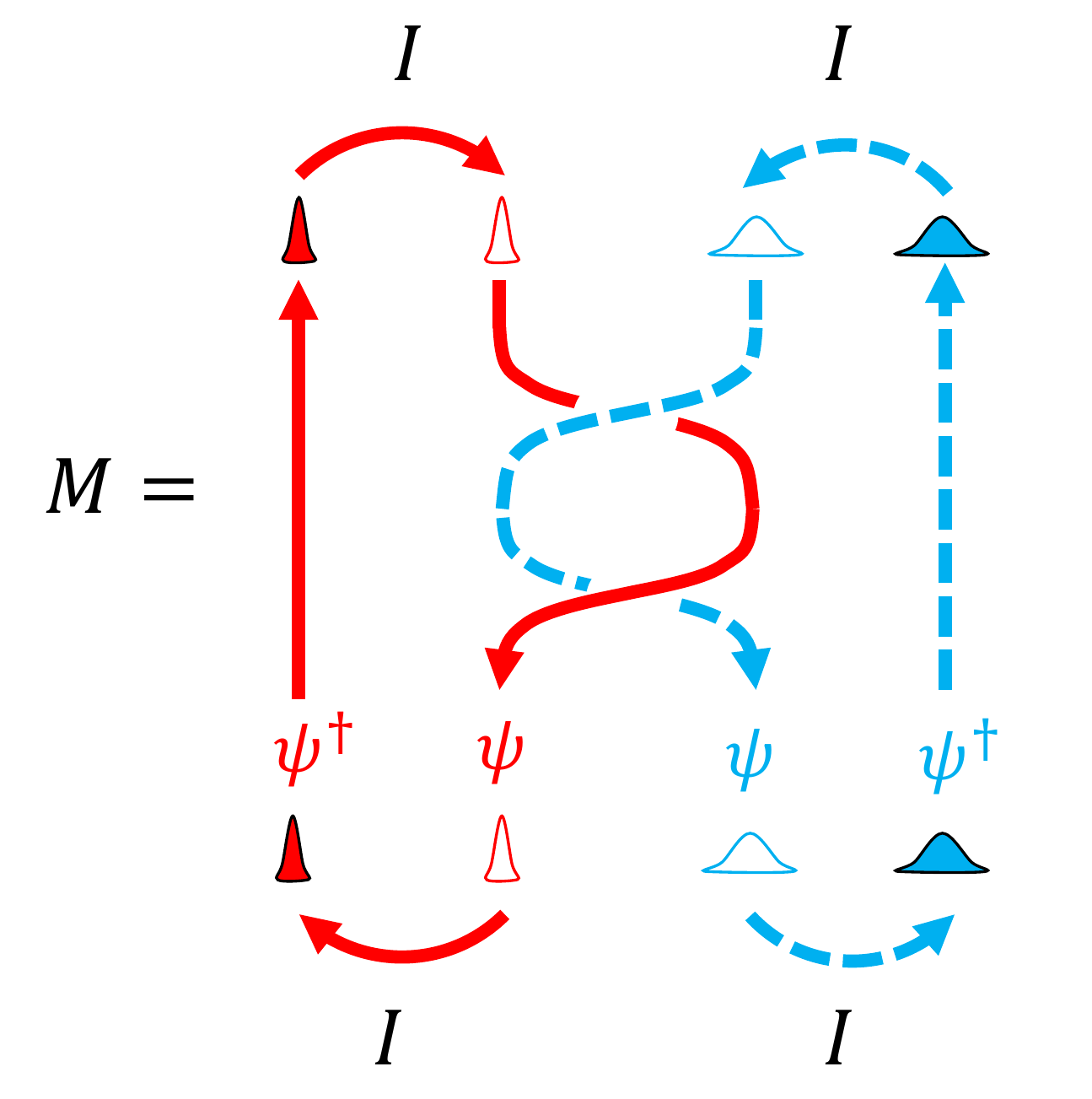}
	\caption{Monodromy $M$ for non-Abelian anyons. Two particle-hole pairs of $\psi$ anyons are initially split from the vacuum. After the braiding, they fuse into the vacuum. The monodromy is the amplitude of this process.}\label{fig-monodromy}
\end{figure}

\begin{figure}[t]
	\centering
	\includegraphics[width = 0.45\textwidth]{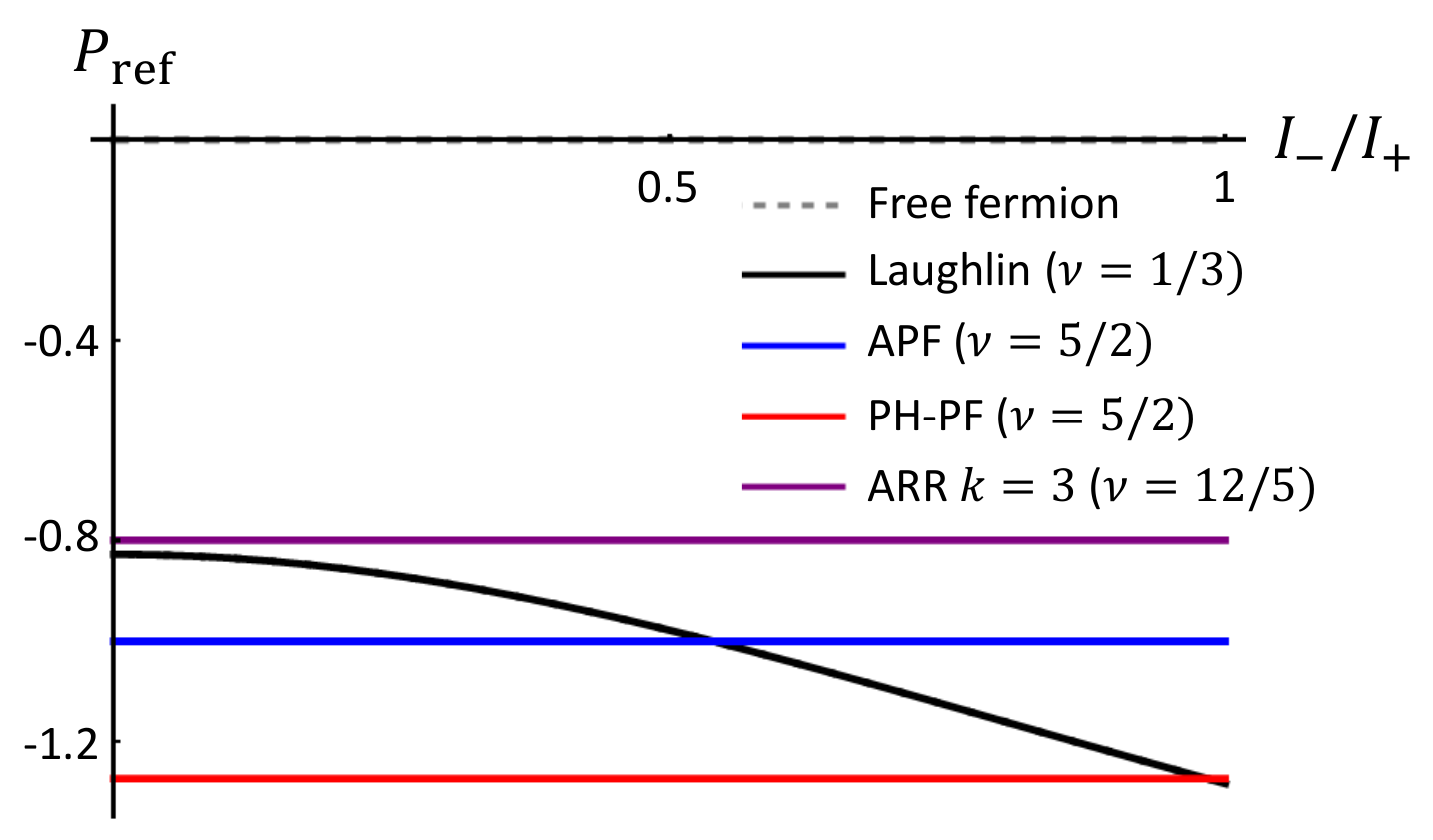}
	\caption{Dependence of Fano factor $P_\textrm{ref}$ on $I_-/I_+$ for free fermions (gray dashed), Laughlin anyons at $\nu=1/3$ (black), anti-Pfaffian state at $\nu = 5/2$ (APf, blue), particle-hole Pfaffian state at $\nu = 5/2$ (PH-Pf, red), and anti-Read-Rezayi state at $\nu = 12/5$ (ARR, pupple). At any value of $I_-/I_+$, $P_\textrm{ref} = -1$ for  the anti-Pfaffian state, $P_\textrm{ref} = -\pi/4$ for  the particle-hole Pfaffian state, and $P_\textrm{ref} = -5\sqrt{250-110\sqrt{5}}/4\pi \simeq - 0.8$ for  the anti-Read-Rezayi state.  The behaviors of the non-Abelian anyons are distingushed from free fermions with $P_\textrm{ref} =0$ and the Abelian anyons.  
}\label{fig2-data}
\end{figure}

We focus on the injection of upstream non-Abelian anyons to QPC$_\textrm{C}$. %
Non-Abelian anyons at $\nu = 5/2$ and 12/5 have $\textrm{Im} [M]=0$ (see their monodromy in Fig.~\ref{fig-monodromy}).
As a notable result, the time-domain interference contributes to the current $I_\textrm{T}$ destructively, 
and Fano factor $P_- (0)$ diverges; the divergence is regularized, $P_- \sim O \big( (\frac{(e^*)^2}{\hbar} \frac{V_\textrm{A/B,inj}}{I_\textrm{A/B,inj}})^{h_a} \big)$, by the subleading terms in Eq.~\eqref{generalI}, where $h_a$ is the scaling dimension of a fusion channel different from the vacuum (Supplementary Information).
For quantitative comparison among anyons, we suggest another Fano factor 
$P_\text{ref}(I_-/I_+) \equiv  (e^*e/h)  \expval{\delta I_\textrm{A} \delta I_\textrm{B}}/(I_+ \partial I_\textrm{T} / \partial V_\textrm{ref}|_{I_-=0, V_\textrm{ref}=0})$ defined with a small reference voltage $V_\textrm{ref}$ applied to Source S$_\textrm{A}'$ and voltage shift $V_\textrm{A,inj} \to V_\textrm{A,inj} + V_\textrm{ref}$ at Source S$_\textrm{A}$ (the voltage across QPC$_\textrm{A}$ remains as $V_\textrm{A,inj}$). 
We find $P_\textrm{ref} (I_-/I_+) = P_-(I_-/I_+) \textrm{Im}[1-M] e /(2 \pi e^*)$. When $I_\text{A,inj} = I_\text{B,inj}$, 
\begin{equation}\label{newfano}
\begin{split}
P_\textrm{ref} (0) =  \frac{\Im[1-M]}{2\pi e^*/e} - \frac{\Re[1-M]}{2\pi e^*/e}	\frac{\cot \pi \delta }{1-2\delta}.	
\end{split}
\end{equation} 
$P_\textrm{ref}$ is notably independent of $I_-/I_+$ for the non-Abelian anyons having $\Im[M] =0$.
In Fig.~\ref{fig2-data}, the behavior of $P_\textrm{ref}$ distinguishes various anyons.
$P_\textrm{ref}$ also differs between the anti-Pfaffian state and the particle-hole Pfaffian state at $\nu = 5/2$;
the states have $M=0$ in common but different $\delta$.
$P_\textrm{ref}$ is experimentally measurable (Methods). It is also possible to gain monodromy information from $I_\textrm{T}$ without measuring $\expval{\delta I_\textrm{A} \delta I_\textrm{B}}$ (Methods).

We compare the time-domain interference with a Fabry-Perot interference~\cite{Chamon97,Nakamura20, Ofek10, Keyserlingk, Willett09, Bishara08,An11, Rosenow09, Bishara09}.
In the latter, an anyon moving around the edge of a Fabry-Perot cavity braids localized bulk anyons inside the cavity. It is detected in linear response of the interference current, with changing the number of the localized anyons by a gate voltage. It corresponds to the interference of free fermions where the braiding is trivial.
By contrast, in the former, braiding happens between anyons on one dimensional edges, as the time ordering provides an extra dimension for braiding. It is detected in the non-equilibrium response, with changing the number of injected anyons by $I_\textrm{A/B,inj}$.

The time-domain interference is absent in free fermions of $M=1$.
For them, the exponential factor in Eq.~\eqref{neqtunneling} becomes the trivial value 1, and the leading contributions in Eq.~\eqref{generalI} vanish.
It is because the topological link between the blue and red loops in Fig.~\ref{fig1}(c) becomes trivial.
The blue loop is completely independent of the red loops, constituting a disconnected Feynman diagram  (a vacuum bubble)  in the perturbation theory.
This diagram cannot contribute to observables, as its contribution $M$ to the interference is exactly cancelled, $M-1 = 0$, by the trivial value 1 from a partner disconnected diagram, according to the linked cluster theorm. 
By contrast, in Abelian and non-Abelian anyons, the cancellation is only partial, $M - 1 \ne 0$. 
We notice, in every perturbation order, pairwise appearance of a braiding diagram and its partner disconnected diagram resulting in the factor $M-1$ (Supplementary Material). This explains $M-1$ in Eqs.~\eqref{neqtunneling} and \eqref{generalI}.
As the time-domain interference has no counterpart in free fermions,
the result in Fig.~\ref{fig2-data} cannot be interpreted as deviation from fermionic antibunching due to the direct collision.
 
We discuss experimental observability.
To observe the Fano factor $P_-$, phase coherence of Edge A/B is required near QPC$_\textrm{C}$ over a distance longer than the thermal length $\hbar v / (k_B T)$, where $v$ is the anyon velocity.
Edge reconstruction \cite{Rosenow02} needs to be avoided over the distance, as it modifies $M$ and $\delta$.
It is also required that QPC$_\textrm{C}$ follows the power law $I \sim V^{2 \delta -1}$ in an energy window which covers the voltages $e^*V_\textrm{A/B,inj}$ and temperature $k_B T$.
The requirements may be achieved in experiments~\cite{Bartolomei20}.
When the energy window also includes the small voltage $e^* V_\textrm{ref}$, the Fano factor $P_\textrm{ref}$ can be measured.
Note that the bulk-edge coupling of non-Abelian anyons~\cite{Bishara09, Rosenow09} and Coulomb interaction~\cite{Keyserlingk} of a Fabry-Perot cavity may be irrelevant in the collider.

It is interesting that the braiding effect appears and dominates the observables in the collider. It differs from the conventional collision, and provides a tool for directly identifying braiding of various Abelian and non-Abelian anyons.
Our finding implies that recent collider experiments~\cite{Bartolomei20} in fact provide a direct signature of Abelian anyon braiding, more direct than the (anti)bunching effects commonly recognized by the community.
Our theory is applicable to other topological orders, as it is based on the generic CFT.
The time-domain interference will be useful for identifying fractional statistics in systems having no topological order~\cite{Jlee20, Morel21} and for engineering of mobile anyons with tuning edge channels by electrical gates.

\begin{center}
	\textbf{METHODS}
\end{center}

\textbf{Tunneling current and noise --- }
We provide the expression of the electrical current $I_\textrm{T}$ and its zero-frequency noise $\expval{\delta I_\textrm{T}^2}$ at QPC$_\text{C}$ at temperature $k_B T \ll e^* V_\textrm{A/B,inj}$ and $h I_\text{A/B,inj}/e^* \ll e^* V_\text{A/B,inj}$,
\begin{equation}\label{tunn_cn_method}
\begin{split}
I_\textrm{T}= &	e^* \int_{-\infty}^\infty dt \expval{\comm{\mathcal{T}^\dagger(0)}{\mathcal{T}(t)}}_\text{neq}
=-C e^*( k_BT)^{2\delta - 1} \Im[\frac{\Gamma( \mathcal{I}/2\pi k_BT + \delta )}{\Gamma(\mathcal{I}/2\pi k_BT + 1-\delta)}], \\
\expval{\delta I_\textrm{T}^2} = & e^{*2}\int_{-\infty}^\infty dt \expval{\acomm{\mathcal{T}^\dagger(0)}{\mathcal{T}(t)}}_\text{neq} 
= \frac{C  e^{*2}}{\tan \pi \delta }( k_BT)^{2\delta - 1} \Re[\frac{\Gamma( \mathcal{I}/2\pi k_BT + \delta )}{\Gamma(\mathcal{I}/2\pi k_BT + 1-\delta)}],
\end{split}
\end{equation}
where %
$C = 4(2\pi)^{2\delta-1}\abs{\gamma_\text{C}}^2 \Gamma(1-2\delta)\sin \pi \delta/d_\psi$. 
This is the generalization of the zero-temperature result for Abelian anyons in Ref.~\cite{Rosenow16} 
to Abelian or non-Abelian anyons at finite temperature.

\textbf{ Cross correlation  --- }
The cross correlation  $\expval{\delta I_\textrm{A} \delta I_\textrm{B}}$ is related with $I_\textrm{T}$ and $\expval{\delta I_\textrm{T}^2}$. Using the charge conservation,  we derive the zero-temperature relations of 
\begin{equation} \begin{split}
\expval{\delta I_\textrm{A} \delta I_\textrm{B}}= & - \expval{\delta I_\textrm{T}^2 } + \expval{\delta I_\textrm{A,inj} \delta I_\textrm{T}} - 	\expval{\delta I_\textrm{B,inj} \delta I_\textrm{T}}   + \expval{\delta I_\textrm{A,inj} \delta I_\textrm{B,inj}},
\quad \expval{\delta I_\textrm{A(B),inj} \delta I_\textrm{T}} = e^* I_\textrm{A(B), inj}\frac{\partial I_\textrm{T}}{\partial I_\textrm{A(B),inj}}.
\end{split} \end{equation}
The latter relation is valid when $\textrm{Im}[M] \ne 0$.
In Supplementary Material, the derivation of the relations, $\expval{\delta I_\textrm{A/B,inj} \delta I_\textrm{T}}$, and $\expval{\delta I_\textrm{A,inj} \delta I_\textrm{B,inj}}$ is found, and the case of $\textrm{Im}[M]=0$ is discussed.

\textbf{ Symmetric injection  --- }
In the nearly symmetric injection case of $I_\text{A,inj} \simeq I_\text{B,inj}$ or $I_{+} \gg I_{-}$, %
the zero-temperature expressions of $I_\textrm{T}$ and $\expval{\delta I_\textrm{T}^2}$ at QPC$_\text{C}$ in Eq.~\eqref{generalI} are simplified as
\begin{equation}\label{diffcond}
\begin{split}
I_\textrm{T} \simeq &\frac{4\abs{\gamma_\text{C}}^2e^*}{d_\psi \csc \pi \delta } (1-2\delta) \Gamma(1-2\delta) \left(\Re[1-M]\frac{I_+}{e^*}\right)^{2\delta - 2}  \left(\frac{e^* V_\textrm{ref}}{\hbar}-\Im[M]\frac{I_-}{e^*}  \right),   \\
\expval{\delta I_\textrm{T}^2} \simeq &\frac{4\abs{\gamma_\text{C}}^2e^{*2}}{d_\psi \sec \pi \delta } \Gamma(1-2\delta) \left(\Re[1-M]\frac{I_+}{e^*}\right)^{2\delta -1}.
\end{split}
\end{equation}
Here we consider the situation where the voltage $V_\textrm{A,inj} + V_\textrm{ref}$ is applied at Source S$_\textrm{A}$, $V_\textrm{B,inj}$ is at Source S$_\textrm{B}$, and a very small voltage $V_\textrm{ref}$ is at Source S$_\textrm{A}'$. In this situation,  the voltage cross QPC$_\textrm{A}$ remains as $V_\textrm{A,inj}$. The effect of $V_\textrm{ref}$  does not modify Eq.~\eqref{generalI} except the replacement of  $\mathcal{I}  \to   \mathcal{I} = \Re[1-M]  \frac{I_+}{ e^*} + i \Im[1-M] \frac{I_-}{  e^*}  + i \frac{e^*}{\hbar} V_\textrm{ref}$. $V_\textrm{ref}$ decouples from the monodromy factor $(1-M)$ in $\mathcal{I}$, as it does not cause any braiding.

\textbf{ Properties of non-Abelian anyons --- }
We briefly intoduce the anti-Read-Rezayi (ARR) state at the level-$k$, a promising candidate hosting non-Abelian anyonic excitations \cite{Bishara08-ARR}.
It has been expected that it is the ground state at $\nu = 2+ \frac{2}{k+2}$. 
In particular, the ARR states of level 2 and of level 3 correspond to the anti-Pfaffian %
state at $\nu = 5/2$ and the ARR state at $\nu = 12/5$, respectively.
The edge-channel structure of the level-$k$ ARR state is decomposed, as a result of random inter-edge tunneling~\cite{  Bishara08-ARR, Lee07, Levin07}, into downstream charge modes, described by the free boson CFT, and an upstream neutral mode, described by the $\text{SU}(2)_k$ Wess-Zumino-Witten CFT. There are two types of the quasiparticles with the smallest scaling dimension of $ h_\psi = 1/(k+2)$ and hence the smallest tunneling exponent $\delta  = 2h_\psi = 2/(k+2)$ for ideal edges, one carrying only charge $e^* = 2e/(k+2)$, and the other carrying $e^* = e/(k+2)$ and the neutral part $j = 1/2$ in the context of the $\text{SU}(2)_k$ anyons. As the bare tunneling strength of the former at a QPC is expected to be much smaller than the latter, we assume that tunneling at the QPCs is dominated by the latter having non-Abelian anyons in the neutral part. The monodromy of the non-Abelian anyons is $M = \frac{\cos(2\pi/(k+2))}{\cos(\pi/(k+2))}$.

We also consider the particle-hole symmetric Pfaffian %
state, another competitive ground state candidate of $\nu = 5/2$~\cite{Banerjee18}. Its edge structure is similar to the anti-Pfaffian state, except that the neutral mode is described by the Ising CFT, and charge $e/4$ quasiparticle contains the non-Abelian anyonic $\sigma$ primary field with scaling dimension of $1/8$ \cite{Zucker}. The monodromy of the non-Abelian anyon is $M=0$.
 
\textbf{ Differential conductances --- } 
We suggest how $\partial I_\textrm{T}/ \partial I_-$ and 
$\partial I_\textrm{T} / \partial V_\textrm{ref}$, hence, the Fano factors $P_-$ and $P_\textrm{ref}$,
can be obtained from standard lock-in measurements. 
First, to obtain $\partial I_\textrm{T}/ \partial I_-$, one applies a small AC voltage to Source S$_\textrm{A}$ in the presence of the voltages $V_\textrm{A/B,inj}$ at QPC$_\textrm{A/B}$, and measures the AC current at Detector D$_\textrm{B}$. Then one gets the differential conductance of
\begin{equation}\label{diffcond_measure1}
\frac{dI_{\text{D}_\text{B}}^{(1)} }{dV}= \frac{dI_\textrm{T}}{dV} = \frac{\partial I_\textrm{T} }{\partial I_\text{A,inj}} \frac{\partial I_\text{A,inj} }{\partial V_\textrm{A,inj}}= G T_\text{A} \frac{\partial I_\textrm{T} }{\partial I_\text{A,inj}}.
\end{equation}
$G$ is  the conductance quantum $e^* e /h$.
$T_\text{A} \equiv G^{-1}\partial I_\text{A,inj}/\partial V_\text{A,inj}$ is the transmission probability at QPC$_\text{A}$, and it can be measured by another lock-in measurement.
From this, one can obtain $\partial I_\textrm{T} /\partial I_\text{A,inj}$. In the limit of $I_- =0$, $\partial I_\textrm{T}/ \partial I_-$ is identical to $\partial I_\textrm{T} /\partial I_\text{A,inj}$.
At nonzero $I_-$, one has a similar measurement for $\partial I_\textrm{T} /\partial I_\text{B,inj}$, and obtains $\partial I_\textrm{T}/ \partial I_- = (\partial I_\textrm{T}/\partial I_\text{A,inj} - \partial I_\textrm{T}/\partial I_\text{B,inj})/2$.

Next, to obtain $\partial I_\textrm{T} / \partial V_\textrm{ref}$, 
one applies a small AC voltage to Source S$_\textrm{A}'$ in the presence of the voltages $V_\textrm{A/B,inj}$ at QPC$_\textrm{A/B}$, and measures the AC current at Detector D$_\textrm{B}$. Then one gets the differential conductance of
\begin{equation}\label{diffcond_measure2}
\frac{dI_{\text{D}_\text{B}}^{(2)} }{dV}= \frac{dI_\textrm{T}}{dV} = - \frac{\partial I_\textrm{T} }{\partial I_\text{A,inj}} \frac{\partial I_\text{A,inj} }{\partial V_\textrm{A,inj}} +\frac{\partial I_\textrm{T} }{\partial V_\text{ref}} = - GT_\text{A}\frac{\partial I_\textrm{T} }{\partial I_\text{A,inj}}  +\frac{\partial I_\textrm{T} }{\partial V_\text{ref}}.
\end{equation}
Combining $dI_{\text{D}_\text{B}}^{(1)}/dV$ and $dI_{\text{D}_\text{B}}^{(2)}/dV$, one can obtain $\partial I_\textrm{T} / \partial V_\textrm{ref}$. 
Note that in the equality in Eq.~\eqref{diffcond_measure2}, $I_\textrm{A,inj}$ and $V_\text{ref}$ are treated as independent variables.
It is because we consider the situation of the voltage $V_\textrm{A,inj} + V_\textrm{ref}$ applied at Source S$_\textrm{A}$, $V_\textrm{B,inj}$ at Source S$_\textrm{B}$, and a very small voltage $V_\textrm{ref}$ at Source S$_\textrm{A}'$; in this situation, the voltage across the QPC$_\textrm{A}$ (hence $I_\textrm{A,inj}$) is independent of $V_\textrm{ref}$.

It is possible to gain monodromy information from the differential conductances without measuring the cross correlation,
since the time-domain interference involving the braiding affects the tunneling current $I_\textrm{T}$. From Eqs.~\eqref{generalI} and \eqref{diffcond},  we find that the ratio of the differential conductances depends only on the fractional charge and $\textrm{Im}[M]$,
\begin{equation} \label{ratio_conductance2}
\frac{\partial I_\textrm{T}/\partial I_-|_{V_\textrm{ref}=0}}{\partial I_\textrm{T} / \partial V_\textrm{ref}|_{V_\textrm{ref}=0}}=\frac{\hbar}{(e^*)^2} \Im[1-M].
\end{equation}
Interestingly, this ratio is independent of $I_+$ and $I_-$. For those non-Abelian anyons having $\textrm{Im}[M]=0$, this ratio shows a vanishingly small value of  $O \big( (\frac{\hbar}{(e^*)^2} \frac{I_\textrm{A/B,inj}}{V_\textrm{A/B,inj}})^{h_a} \big)$.  
The ratio can be measured when QPC$_\textrm{C}$ follows the power law $I \sim V^{2 \delta -1}$ in an energy window which covers the voltages $e^*V_\textrm{A/B,inj}$, temperature $k_B T$, and small voltage $e^* V_\textrm{ref}$.

\begin{center}
	\textbf{ACKNOWLEDGEMENTS}
\end{center}

We thank Anne Anthore, Hyung Kook Choi, Gwendal Feve, Christian Glattli, Christophe Mora, and Frederic Pierre  for useful discussions. This work is supported by Korea NRF via the SRC Center for Quantum Coherence in Condensed Matter (GrantNo.2016R1A5A1008184) and NRF-2019-Global Ph.D. fellowship.

\newpage

\begin{center}
\textbf{\large Supplemental Materials: Non-Abelian Anyon Collider}
\linebreak \\
June-Young M. Lee and H.-S. Sim \\

\textit{\small Department of Physics, Korea Advanced Institute of Science and Technology, Daejeon 34141, Korea}
\end{center}

\setcounter{equation}{0}
\setcounter{figure}{0}
\setcounter{table}{0}
\setcounter{page}{1}
\makeatletter
\renewcommand{\thesection}{S\Roman{section}}
\renewcommand{\theequation}{S\arabic{equation}}
\renewcommand{\thefigure}{S\arabic{figure}}
\renewcommand{\bibnumfmt}[1]{[S#1]}
\renewcommand{\citenumfont}[1]{S#1}

\section{Derivation of the nonequilibrium correlator in Eq. (1)}\label{sec-dercorr}

The non-equilibrium correlator in Eq.~\eqref{neqtunneling} is decomposed into a product of correlators of the edges A and B, $\expval{\mathcal{T}^\dagger(0)\mathcal{T}(t)}_\text{neq} \propto \expval{[\psi^\dagger_\textrm{A}(0,0)\psi_\textrm{A}(0,t)]_I}_\text{neq} \expval{[\psi_\textrm{B}(0,0)\psi_\textrm{B}^\dagger(0,t)]_I}_\text{neq}$. We here derive the correlator of Edge A. %

The tunneling current $I_\text{A,inj}$ is generated at QPC$_\textrm{A}$ from Edge A$_\textrm{inj}$ (the edge directly connected to Source S$_\textrm{A}$) to Edge A by the voltage $V_\text{A,inj}$ applied to Source S$_\textrm{A}$.
We assume that the weak coupling regime is dominated by tunneling of a single quasiparticle type,
described by the tunneling Hamiltonian $H_\text{A,inj} = \mathcal{A}(t) + \mathcal{A}^\dagger(t)$,
\begin{eqnarray} \label{full_tunneling_H}
\mathcal{A}(t)= \gamma_\text{A} e^{-ie^*V_\text{A,inj}t/\hbar} [\psi_\text{A,inj}(0,t) \psi^\dagger_\textrm{A}(-d,t)]_I \cdots.
\end{eqnarray}
with the tunneling strength $\gamma_\text{A}$ and the charge  $e^*$ carried by the quasiparticle. 
At certain filling factors, the quasiparticle is fractionalized into downstream and upstream parts upon its tunneling.
We focus on the fractionalized part $\psi_\textrm{A}$ moving from QPC$_\textrm{A}$ to QPC$_\textrm{C}$ along Edge A, since 
this determines $I_\textrm{T}$, $\langle \delta I_\textrm{T}^2 \rangle$, $\langle \delta I_\textrm{A} \delta I_\textrm{B} \rangle$.
 $\psi_\textrm{A,inj}$ is the corresponding part on Edge A$_\textrm{inj}$. $[ \cdots ]_{a=I}$ means that $\psi_\textrm{A}^\dagger$ and $\psi_\textrm{A,inj}$ are pair-created from the vacuum fusion channel $a=I$.
The other part has negligible contribution to the observables (see Sec.~\ref{sec-back}),
and omitted as $\cdots$ in Eq.~\eqref{full_tunneling_H}. The anyon $\psi_\textrm{A}$ is described by the primary field of the relevant conformal field theory (CFT) with the scaling dimension $h_\psi$ and quantum dimension $d_\psi$. 
We will derive the correlator of the anyons on Edge A at the voltage $V_\text{A,inj}$,
\begin{equation}
\begin{split}\label{neqcorr}
\expval{[\psi_\textrm{A}(x,t)\psi_\textrm{A}^\dagger(0,0)]_I}_\text{neq}  = e^{\Re[M-1]\frac{I_\text{A,inj}}{e^*}\abs{t-\frac{x}{v}}}e^{i\Im[M-1]\frac{I_\text{A,inj}}{e^*}(t-\frac{x}{v})}\expval{[\psi_\textrm{A}(x,t)\psi_\textrm{A}^\dagger(0,0)]_I}_\text{eq} + \textrm{subleading terms}.
\end{split}
\end{equation}
$\expval{[\psi_\textrm{A}\psi_\textrm{A}^\dagger]_I}_\text{eq}$ is the equilibrium correlator at $V_\text{A,inj} =0 $. $v$ is the anyon velocity.
We note that the  tunneling current is expressed as $I_\text{A,inj}= 2\pi \abs{\gamma_\text{A}}^2 e^*(e^*V_\text{A,inj}/\hbar)^{4h_\psi-1}/[d_\psi\Gamma(4h_\psi)]$ when the quasiparticle is fully described by $\psi$ (i.e., not fractionalized).
Below we drop the edge index A and focus on $vt-x >0$; the  $vt-x < 0$ case is obtained similarly. 

We perform the Keldysh pertubative expansion $\expval{[\psi(x,t)\psi^\dagger(0,0)]_I}_\text{neq} = \sum_{n=0}^\infty \mathcal{C}_{n}(x,t)$ over arbitrary orders of the tunneling strength at QPC$_\textrm{A}$. 
$\mathcal{C}_{n}(x,t)$ is the $2n$-th order perturbation term proportional to $\abs{\gamma_\text{A}}^{2n}$ and $(-i)^{2n}/(2n)!$,
\begin{equation}
\begin{split}
&\mathcal{C}_{n}(x,t)= \frac{(-1)^{n}}{(2n)!}\frac{(2n)!}{n!n!}  \int_K dt_1\cdots dt_{2n} \langle T_K\big\{[\psi(x,t)\psi^\dagger(0,0)]_I \prod_{i=1}^n \mathcal{A}^\dagger(t_{2i-1})\mathcal{A}(t_{2i})\big\}\rangle.			 
\end{split}
\end{equation}
We set $\hbar \equiv 1$. $T_K\{\cdots\}$ is the Keldysh time ordering. $\int_K dt_1\cdots dt_{2n}$ is the time $t_i$ integration on the Keldysh contour.  $\prod_{i=1}^n \mathcal{A}^\dagger(t_{2i-1})\mathcal{A}(t_{2i})$ is a combination of $n$ $\mathcal{A}$ and $n$ $\mathcal{A}^\dagger$ chosen from $2n$ $H_\text{inj}$ operators. %
The total number of such combinations is $(2n)!/n!n!$. 
It is obvious that $\mathcal{C}_0(x,t)$ equals the equilibrium correlator $\expval{[\psi(x,t)\psi^\dagger(0,0)]_I}_\text{eq}$.

We compute $\mathcal{C}_{n}$ in the regime of large $e^*V_\text{inj}$.
The integral in $\mathcal{C}_{n}$ is dominated by the time windows of $t_i$'s where 
anyons from $n$ $\mathcal{A}^\dagger(t_{2i-1})$'s and those from $n$ $\mathcal{A}(t_{2i})$'s pairwise overlap on Edge A, 
i.e., the anyons in each pair are located within short distance $\hbar v / (e^* V_\text{inj})$;
when $e^*V_\text{inj}$ is the largest energy, it is a good approximation to set $t_{2i - 1} \simeq t_{2i'}$ (pairing time indices $2i-1$ and $2i'$).
We set $t_{2i-1} \simeq t_{2i}$ and order them as $t_{1} \simeq t_2 < t_3 \simeq t_4 < \cdots$, without loss of the generality,
with multiplying the number $n!$ of the equivalent ways of the pairings to the integral. The Keldysh index at time $t_j$ is $\eta_j = + 1$ for the forward branch of the Keldysh contour and $\eta_j = - 1$ for the backward branch. Then
\begin{equation}
\begin{split}\label{introF}
\mathcal{C}_n(x,t) \simeq  &\abs{\gamma_\text{A}}^{2n} \frac{(-1)^n}{n!} \sum_{\eta_j = \pm}\prod_j\eta_j  \int\limits_{t_{2i-1} \simeq t_{2i}} dt_1 \cdots dt_{2n} e^{i\omega\sum_i (t_{2i-1}-t_{2i})}\mathcal{F}^{\vec{\eta}}, \end{split}
\end{equation}
where $\omega = e^* V_\textrm{inj} / \hbar$.
$\mathcal{F}^{\vec{\eta}}$ is the conformal block of the primary fields $\psi$'s,
\begin{equation} \label{conformal_block}
\begin{split}
\mathcal{F}^{\vec{\eta}}\equiv &	\frac{e^{-i\omega \sum_i(t_{2i-1}-t_{2i})}}{\abs{\gamma_\text{A}}^{2n}}	\langle T_K\big\{[\psi(x,t)\psi^\dagger(0,0)]_I \prod_{i=1}^n \mathcal{A}^\dagger(t^{\eta_{2i-1}}_{2i-1})\mathcal{A}(t^{\eta_{2i}}_{2i})\big\}\rangle.		\\
\end{split}
\end{equation}
Here the factor $e^{-i\omega \sum_i(t_{2i-1}-t_{2i})} / \abs{\gamma_\text{A}}^{2n}$ cancels the corresponding factors in $\mathcal{A}$'s so that $\mathcal{F}^{\vec{\eta}}$ contains only the anyon fields. In the conformal block $\mathcal{F}^{\vec{\eta}}$, $\psi (x,t)$ and $\psi^\dagger(0,0)$ are the main fields of the correlator, while the others $\psi(-d, t^{\eta_{2i-1}}_{2i-1})$'s and $\psi^\dagger(-d, t^{\eta_{2i}}_{2i})$'s in $\mathcal{A}^\dagger$'s and $\mathcal{A}$'s describe anyons injected to Edge A through QPC$_\textrm{A}$ 
by $V_\textrm{A,inj}$.

For the injected anyons to be correlated with the anyons described by the main fields $\psi (x,t)$ and $\psi^\dagger(0,0)$,
they should be injected before the time $0$ or $t$. Then, after the Keldysh ordering, the conformal block has the ordering of  
\begin{equation}
\begin{split}
\mathcal{F}^{\vec{\eta}} \propto \langle (\text{$\mathcal{A}$, $\mathcal{A}^\dagger$s with $\eta_j = -1$}) [\psi(x,t)\psi^\dagger(0,0)]_I(\text{$\mathcal{A}$, $\mathcal{A}^\dagger$s with $\eta_j = 1$}) \rangle. \nonumber
\end{split}
\end{equation}
In the case where $\eta_j = -1$ for all $j$'s, for example,
$\mathcal{F}^{\vec{\eta}}$ has the following form and diagramatic representation,
\begin{align}
\mathcal{F}^{(-1,\cdots,-1)}  = & \langle [ \psi (-d, t_1) \psi^\dagger_\textrm{inj}(0, t_1)]_I [\psi_\textrm{inj}(0, t_2) \psi^\dagger (-d, t_2) ]_I [\psi (-d, t_3) \psi^\dagger_\textrm{inj}(0, t_3) ]_I \cdots [\psi(x,t)\psi^\dagger(0,0)]_I \rangle \nonumber \\
= &\begin{tikzpicture}[baseline=(current  bounding  box.center),declare function = {}]
\node at (0, 0.75) {$d_\psi^{-(n+\frac{1}{2})}$};
\draw[gray, dashed,thick](0.6,0.3)--(7.7,0.3);
\foreach \pos in {1.2} {
	\draw[gray,dashed, thick](\pos,0.3)--(\pos,0.6);
	\draw[-to,black,thick](\pos-0.4,1.0)..controls(\pos-0.4,0.5) and(\pos+0.4,0.5)..(\pos+0.4,1.0);
	\node at (\pos-0.4,1.2) {$1$};
	\node at (\pos+0.4,1.2) {$1'$};
}
\foreach \pos in {2.5} {
	\draw[gray,dashed, thick](\pos,0.3)--(\pos,0.6);
	\draw[-to,black,thick](\pos-0.4,1.0)..controls(\pos-0.4,0.5) and(\pos+0.4,0.5)..(\pos+0.4,1.0);
	\node at (\pos-0.4,1.2) {$2'$};
	\node at (\pos+0.4,1.2) {$2$};
}
\foreach \pos in {3.8} {
	\draw[gray,dashed, thick](\pos,0.3)--(\pos,0.6);
	\draw[-to,black,thick](\pos-0.4,1.0)..controls(\pos-0.4,0.5) and(\pos+0.4,0.5)..(\pos+0.4,1.0);
	\node at (\pos-0.4,1.2) {$3$};
	\node at (\pos+0.4,1.2) {$3'$};
}
\foreach \pos in {5.1} {
	\draw[gray,dashed, thick](\pos,0.3)--(\pos,0.6);
	\draw[-to,black,thick](\pos-0.4,1.0)..controls(\pos-0.4,0.5) and(\pos+0.4,0.5)..(\pos+0.4,1.0);
	\node at (\pos-0.4,1.2) {$4'$};
	\node at (\pos+0.4,1.2) {$4$};
}
\foreach \pos in {7.1} {
	\draw[gray,dashed, thick](\pos,0.3)--(\pos,0.6);
	\draw[-to,black,thick](\pos-0.4,1.0)..controls(\pos-0.4,0.5) and(\pos+0.4,0.5)..(\pos+0.4,1.0);
	\node at (\pos-0.4,1.2) {$\psi$};
	\node at (\pos+0.4,1.2) {$\psi^\dagger$};
}
\foreach \pos in {0.0, 1.2,2.5,3.8,4.9, 6,7}{
	\node at (\pos+0.6,0.1) {\textcolor{gray}{$I$}};
}
\node at (6.1,0.7) {$\cdots$}; 
\end{tikzpicture}\label{F-corr}
\end{align}
where the head (resp. tail) of the arrow with numbers $j$, $j'$ depicts the creation (resp. annihilation) operator of the square braket at $t_j$. The numbers without (with) the prime symbol depict anyons on Edge A (A$_\textrm{inj}$). The symbol $I$ means the vacuum fusion channel. %
We use the normalization convention of the diagrams in Ref.~\cite{Bonderson-thesis-S}. %

For further computation, we express the conformal block $\mathcal{F}^{\vec{\eta}}$ as a linear combination~\cite{Fendley07-S} of the fusion bases $\mathcal{G}^{\vec{\eta}}_{\vec{a},\vec{b}}$'s in which the fields at similar spacetime $t_{2i-1} \simeq t_{2i}$ on the same edge fuse first to an intermediate fusion state $a_i$,
\begin{align}
\mathcal{G}^{\vec{\eta}}_{\vec{a},\vec{b}}
= & \expval{[[[[[\psi(-d,t_1)\psi^\dagger(-d,t_2)]_{a_1=b_1}[\psi^\dagger_\text{inj}(0,t_1)\psi_\text{inj}(0,t_2)]_{a_2}]_{b_2}[\psi(-d,t_3)\psi^\dagger(-d,t_4)]_{a_3}]_{b_3}\cdots ]_{b_{2n}}[\psi(x,t)\psi^\dagger(0,0)]_{a_{2n+1}}]_{I}} \nonumber \\
= & \begin{tikzpicture}[baseline=(current  bounding  box.center),declare function = {}]
\node at (0, 0.75) {$d_\psi^{-(n+\frac{1}{2})}$};
\draw[black,thick](0.6,0.3)--(7.7,0.3);
\foreach \pos in {1.2} {
	\draw[black, thick](\pos,0.3)--(\pos,0.6);
	\draw[-to,black,thick](\pos-0.4,1.0)..controls(\pos-0.4,0.5) and(\pos+0.4,0.5)..(\pos+0.4,1.0);
	\node at (\pos-0.4,1.2) {$1$};
	\node at (\pos+0.4,1.2) {$2$};
}
\foreach \pos in {2.5} {
	\draw[black, thick](\pos,0.3)--(\pos,0.6);
	\draw[to-,black,thick](\pos-0.4,1.0)..controls(\pos-0.4,0.5) and(\pos+0.4,0.5)..(\pos+0.4,1.0);
	\node at (\pos-0.4,1.2) {$1'$};
	\node at (\pos+0.4,1.2) {$2'$};
}
\foreach \pos in {3.8} {
	\draw[black, thick](\pos,0.3)--(\pos,0.6);
	\draw[-to,black,thick](\pos-0.4,1.0)..controls(\pos-0.4,0.5) and(\pos+0.4,0.5)..(\pos+0.4,1.0);
	\node at (\pos-0.4,1.2) {$3$};
	\node at (\pos+0.4,1.2) {$4$};
}
\foreach \pos in {5.1} {
	\draw[black, thick](\pos,0.3)--(\pos,0.6);
	\draw[to-,black,thick](\pos-0.4,1.0)..controls(\pos-0.4,0.5) and(\pos+0.4,0.5)..(\pos+0.4,1.0);
	\node at (\pos-0.4,1.2) {$3'$};
	\node at (\pos+0.4,1.2) {$4'$};
}
\foreach \pos in {7.1} {
	\draw[black, thick](\pos,0.3)--(\pos,0.6);
	\draw[-to,black,thick](\pos-0.4,1.0)..controls(\pos-0.4,0.5) and(\pos+0.4,0.5)..(\pos+0.4,1.0);
	\node at (\pos-0.4,1.2) {$\psi$};
	\node at (\pos+0.4,1.2) {$\psi^\dagger$};
}
\foreach \pos in {1,2,3,4}{
	\node at (\pos*1.3+0.15,0.45) {$a$\textsubscript{\pos}};
}
\foreach \pos in {1,2,3}{
	\node at (\pos*1.3+0.7,0.1) {$b$\textsubscript{\pos}};
}
\node at (5.9*1.3,0.45) {$a_{2n+1}$};
\node at (5.6,0.1) {$b_4$};
\node at (6.7,0.1) {$b_{2n}$};
\node at (0.6,0.1){$I$};
\node at (7.6,0.1){$I$};
\node at (6.1,0.7) {$\cdots$};
\end{tikzpicture}. \label{G-corr}
\end{align}
$[[\cdots]_{b_j} [\cdots]_{a_{j+1}}]_{b_{j+1}}$ means the fusion of intermediate states $b_j$ and $a_{j+1}$ into $b_{j+1}$.  The operator product expansion~\cite{Francesco-S}  (OPE) is applied at $t_{2i-1}\simeq t_{2i}$, to obtain $\mathcal{G}^{\vec{\eta}}_{\vec{a},\vec{b}} \propto (t_{2i-1} - t_{2i})^{h_{a_{2i-1}} + h_{a_{2i}}-4h_\psi}$. $h_{a}$ is the scaling dimension of the primary field $a$.  At large $V_\text{inj}$, $\mathcal{F}^{\vec{\eta}}$ is determined by the basis $\mathcal{G}^{\vec{\eta}}_{\vec{I},\vec{I}}$ with $a_j = b_j = I$   where all the intermediate states are the vacuum field of the lowest scaling dimension $h_I=0$. 
One has $\mathcal{G}^{\vec{\eta}}_{\vec{I},\vec{I}} \simeq \langle[\psi_\textrm{A}(x,t)\psi_\textrm{A}^\dagger(0,0)]_I\rangle_\text{eq}\prod_i [\epsilon +i\chi_{\eta_{2i-1},\eta_{2i}}(t_{2i-1} - t_{2i})]^{-4h_\psi}$ with $\chi_{\eta_{2i-1},\eta_{2i}} = [(\eta_{2i-1}+\eta_{2i})\text{sgn}(t_{2i-1} - t_{2i})- (\eta_{2i-1}-\eta_{2i})]/2$. $\epsilon$ is a positive infinitesimal cutoff.

To compute the overlap of $\mathcal{F}^{\vec{\eta}}$ and $\mathcal{G}^{\vec{\eta}}_{\vec{I},\vec{I}}$, we glue the  diagrams $\mathcal{F}^{\vec{\eta}}$ and $\mathcal{G}^{\vec{\eta}}_{\vec{I},\vec{I}}$, reversing arrow directions in the diagram $\mathcal{G}^{\vec{\eta}}_{\vec{I},\vec{I}}$ and connecting the arrow of index $j$ of $\mathcal{G}^{\vec{\eta}}_{\vec{I},\vec{I}}$ with the arrow of the same index of $\mathcal{F}^{\vec{\eta}}$. When different connections $j_1$ and $j_2$ cross, {connection $j_1$ is drawn on top of 
$j_2$ if $x_{j_1} - vt_{j_1} > x_{j_2} - vt_{j_2}$  \cite{braidingrule-S}.  
The overlap depends on the Keldysh indices $\eta_i$. 
In the case of $\eta_j = -1$ for all $j$'s, the gluing of  Eqs.~\eqref{F-corr} and  \eqref{G-corr}  has
$n+1$ unlinked loops, 
\def\y{4.3} 
\def\wid{0.56}
\def\xinit{\wid*2+0.2}
\def\marg{0.1}
\begin{align}\label{overlap-1} 
& \frac{1}{d_\psi^{2n+1}}
\begin{tikzpicture}[baseline=(current  bounding  box.center),
declare function =  {
	slope(\xi,\yi,\xj,\yj) = (\yi-\yj)/(\xi-\xj);
	yintercept(\xi,\yi,\xj,\yj) = \yi - \xi* slope(\xi,\yi,\xj,\yj);
	interx(\xi,\yi,\xj,\yj,\xk,\yk,\xl,\yl) = -(yintercept(\xi,\yi,\xj,\yj)-yintercept(\xk,\yk,\xl,\yl))/(slope(\xi,\yi,\xj,\yj)-slope(\xk,\yk,\xl,\yl));
	intery(\xi,\yi,\xj,\yj,\xk,\yk,\xl,\yl) =interx(\xi,\yi,\xj,\yj,\xk,\yk,\xl,\yl)* slope(\xi,\yi,\xj,\yj) +yintercept(\xi,\yi,\xj,\yj);
}]
\draw[gray, dashed,thick](0.0,0.3)--(\wid*12,0.3);
\node at (\xinit - \wid*2,0.7) {$\cdots$};
\foreach \pos in {\xinit} {
	\draw[gray,dashed, thick](\pos,0.3)--(\pos,0.6);
	\draw[-to,black,thick](\pos-\wid,0.5 + \wid*5/4)..controls(\pos-\wid,0.5) and(\pos+\wid,0.5)..(\pos+\wid,0.5 + \wid*5/4);
	\node at (\pos-\wid,0.5 + \wid*1.6) {$2i-1$};
	\node at (\pos+\wid,0.5 + \wid*1.6) {$2i-1'$};
}
\foreach \pos in {\xinit + \wid*4} {
	\draw[gray,dashed, thick](\pos,0.3)--(\pos,0.6);
	\draw[-to,black,thick](\pos-\wid,0.5 + \wid*5/4)..controls(\pos-\wid,0.5) and(\pos+\wid,0.5)..(\pos+\wid,0.5 + \wid*5/4);
	\node at (\pos-\wid,0.5 + \wid*1.6) {$2i'$};
	\node at (\pos+\wid,0.5 + \wid*1.6) {$2i$};
}
\node at (\xinit + \wid*6,0.7) {$\cdots$};
\foreach \pos in {\xinit + \wid*8} {
	\draw[gray,dashed, thick](\pos,0.3)--(\pos,0.6);
	\draw[-to,black,thick](\pos-\wid,0.5 + \wid*5/4)..controls(\pos-\wid,0.5) and(\pos+\wid,0.5)..(\pos+\wid,0.5 + \wid*5/4);
	\node at (\pos-\wid,0.5 + \wid*1.6) {$\psi$};
	\node at (\pos+\wid,0.5 + \wid*1.6) {$\psi^\dagger$};
}
\draw[gray, dashed,thick](0.0,\y-0.3)--(\wid*12,\y-0.3);
\node at (\xinit - \wid*2,\y-0.7) {$\cdots$};
\foreach \pos in {\xinit} {
	\draw[gray,dashed, thick](\pos,\y-0.3)--(\pos,\y-0.6);
	\draw[to-,black,thick](\pos-\wid,\y-0.5 - \wid*5/4)..controls(\pos-\wid,\y-0.5) and(\pos+\wid,\y-0.5)..(\pos+\wid,\y-0.5 - \wid*5/4);
	\node at (\pos-\wid,\y-0.5 - \wid*1.6) {$2i-1$};
	\node at (\pos+\wid,\y-0.5 - \wid*1.6) {$2i$};
}
\foreach \pos in {\xinit+\wid*4} {
	\draw[gray,dashed, thick](\pos,\y-0.3)--(\pos,\y-0.6);
	\draw[-to,black,thick](\pos-\wid,\y-0.5 - \wid*5/4)..controls(\pos-\wid,\y-0.5) and(\pos+\wid,\y-0.5)..(\pos+\wid,\y-0.5 - \wid*5/4);
	\node at (\pos-\wid,\y-0.5 - \wid*1.6) {$2i-1'$};
	\node at (\pos+\wid,\y-0.5 - \wid*1.6) {$2i'$};
}
\foreach \pos in {\xinit+\wid*8} {
	\draw[gray,dashed, thick](\pos,\y-0.3)--(\pos,\y-0.6);
	\draw[to-,black,thick](\pos-\wid,\y-0.5 - \wid*5/4)..controls(\pos-\wid,\y-0.5) and(\pos+\wid,\y-0.5)..(\pos+\wid,\y-0.5 - \wid*5/4);
	\node at (\pos-\wid,\y-0.5 - \wid*1.6) {$\psi$};
	\node at (\pos+\wid,\y-0.5 - \wid*1.6) {$\psi^\dagger$};
}
\node at (\xinit + \wid*6,\y-0.7) {$\cdots$};
\draw[to-,black,thick](\xinit-\wid,0.5 + \wid*2.0)--(\xinit-\wid,\y-0.5 - \wid*2.0);
\draw[-to,black,thick](\xinit+\wid*5,0.5 + \wid*2.0)--(\xinit+\wid,\y-0.5 - \wid*2.0);
\draw[to-,black,thick](\xinit+\wid*7,0.5 + \wid*2.0)--(\xinit+\wid*7,\y-0.5 - \wid*2.0);
\draw[-to,black,thick](\xinit+\wid*9,0.5 + \wid*2.0)--(\xinit+\wid*9,\y-0.5 - \wid*2.0);
\def\Xi{\xinit+\wid*3}; \def\Yi{0.5 + \wid*2.0};\def\Xj{\xinit+\wid*5}; \def\Yj{\y-0.5 - \wid*2.0};
\def\Xk{\xinit+\wid*5}; \def\Yk{0.5 + \wid*2.0};\def\Xl{\xinit+\wid}; \def\Yl{\y-0.5 - \wid*2.0};
\draw[to-,black,thick](\Xi,\Yi)--({interx(\Xi,\Yi,\Xj,\Yj,\Xk,\Yk,\Xl,\Yl)-\marg },{intery(\Xi,\Yi,\Xj,\Yj,\Xk,\Yk,\Xl,\Yl) -\marg*slope(\Xi,\Yi,\Xj,\Yj) });
\draw[black,thick]({interx(\Xi,\Yi,\Xj,\Yj,\Xk,\Yk,\Xl,\Yl)+\marg },{intery(\Xi,\Yi,\Xj,\Yj,\Xk,\Yk,\Xl,\Yl) +\marg*slope(\Xi,\Yi,\Xj,\Yj) })--(\Xj,\Yj);
\def\Xi{\xinit+\wid}; \def\Yi{0.5 + \wid*2.0};\def\Xj{\xinit+\wid*3}; \def\Yj{\y-0.5 - \wid*2.0};
\def\Xk{\xinit+\wid*5}; \def\Yk{0.5 + \wid*2.0};\def\Xl{\xinit+\wid}; \def\Yl{\y-0.5 - \wid*2.0};
\draw[black,thick](\Xi,\Yi)--({interx(\Xi,\Yi,\Xj,\Yj,\Xk,\Yk,\Xl,\Yl)-\marg },{intery(\Xi,\Yi,\Xj,\Yj,\Xk,\Yk,\Xl,\Yl) -\marg*slope(\Xi,\Yi,\Xj,\Yj) });
\draw[-to,black,thick]({interx(\Xi,\Yi,\Xj,\Yj,\Xk,\Yk,\Xl,\Yl)+\marg },{intery(\Xi,\Yi,\Xj,\Yj,\Xk,\Yk,\Xl,\Yl) +\marg*slope(\Xi,\Yi,\Xj,\Yj) })--(\Xj,\Yj);
\end{tikzpicture} 
= \frac{1}{d_\psi^n}.
\end{align}
Each loop is composed of either connections $(2i-1)'$, $(2i)'$, $(2i-1)$ and $(2i)$ describing an (particle-like) anyon on Edge A injected at QPC$_\textrm{A}$ and the remaining (hole-like) anyon on Edge A$_\textrm{inj}$ at time $t_{2i-1} \simeq t_{2i}$,
or connections $\psi (x,t)$ and $\psi^\dagger (0,0)$ describing the main fields of the correlator.
All the loops are unlinked, and each loop equally contributes to the overlap. %
Hence the conformal block $\mathcal{F}^{\vec{\eta}}$ does not involve anyon braiding in the case of $\eta_j = -1$ for all $j$'s.

We next examine the case where $\eta_{2i-1} = \eta_{2i} = 1$ for some $i$. Then in $\mathcal{F}^{\vec{\eta}}$, after the Keldysh ordering, $\mathcal{A}^\dagger(t^+_{2i-1})$ and $\mathcal{A}(t^+_{2i})$ are placed right of the anyon fields having $\eta_j =-1$, the main fields $\psi(x,t)$ and $\psi^\dagger(0,0)$, and the fields having $\eta_{j}=1$ at $t_j > t_{2i-1} \simeq t_{2i}$. No anyon field is placed between $\mathcal{A}^\dagger(t^+_{2i-1})$ and $\mathcal{A}(t^+_{2i})$ since $t_{2i-1}\simeq t_{2i}$. 
In the gluing of $\mathcal{F}^{\vec{\eta}}$ and $\mathcal{G}^{\vec{\eta}}$, the loop composed of connections $\mathcal{A}^\dagger(t^+_{2i-1})$ and $\mathcal{A}(t^+_{2i})$ is trivially unlinked from all the other loops, %
because $\mathcal{A}^\dagger(t^+_{2i-1})$ and $\mathcal{A}(t^+_{2i})$ always together cross some other connections so that the crossings are trivially loosened. 
Similarly, any loop is trivially unlinked from all the others. No braiding happens among the injected anyons.

\begin{figure}[b]
	\centering
	\includegraphics[width = 0.9\textwidth]{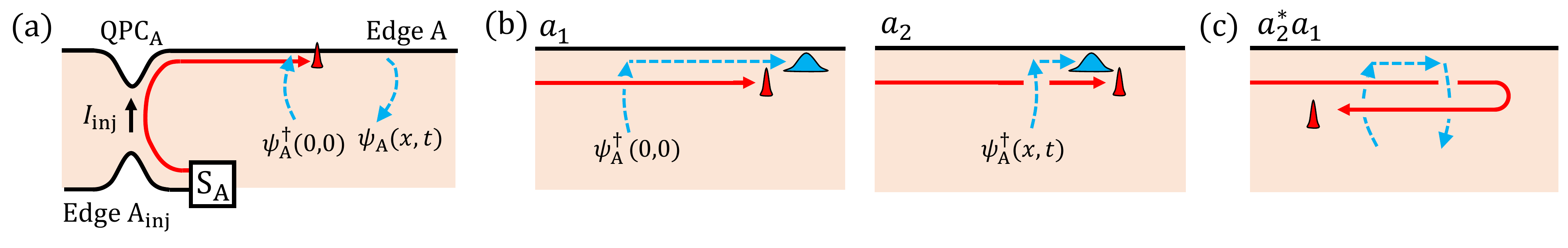}
	\caption{A process contributing to $\mathcal{C}_1(x,t)$.  (a) An anyon is injected from Edge A$_\textrm{inj}$ to Edge A by tunneling at QPC$_\textrm{A}$.  This is described by $\mathcal{A}^\dagger(t_{2i-1})$ and $\mathcal{A}(t_{2i})$ in Eq.~\eqref{overlap-2}. As biased by the large voltage $V_\text{inj}$, the anyon has spatial uncertainty $\hbar v / (e^* V_\text{inj})$, hence, $t_{2i-1} \simeq t_{2i}$; its spatial uncertainty and propagation are depicted by red peaks and red solid arrows. Creation and annihilation of another anyon by  the main fields $\psi^\dagger_\textrm{A}(0,0)$ and $\psi_\textrm{A}(x,t)$ of the correlator are shown by blue dashed arrows.  We consider $0 <vt_{2i-1} +d \simeq vt_{2i} +d < vt-x$ and $x,0 > -d$; $x = -d$ is the location of QPC$_\textrm{A}$ on Edge A.
(b) This process is the interference $a_2^* a_1$ of two subprocesses $a_1$ and $a_2$.  In $a_1$   the blue anyon [described by $\psi^\dagger_\textrm{A}(0,0)$] is created at position 0 and time 0 before the red anyon [by $\mathcal{A}^\dagger(t_{2i-1})$] passes the position. In $a_2$, the blue anyon [by $\psi^\dagger_\textrm{A}(x,t)$] is created at $x$ and $t$ after the red anyon  [by $\mathcal{A}^\dagger(t_{2i})$] passes $x$. 
(c) The interference $a_2^* a_1$ is described by merging $a_1$ and the time reversal of $a_2$. It shows the red loop and the blue loop linked by braiding. This corresponds to the two linked loops of the overlap diagram in Eq.~\eqref{overlap-2}. }\label{fig-corr}
\end{figure}

When $\eta_{2i-1} = -\eta_{2i}$ for some $i$ and $0 <vt_{2i-1} +d \simeq vt_{2i} +d < vt-x$, nontrivial links by braiding happen in the overlap between $\mathcal{F}^{\vec{\eta}}$ and $\mathcal{G}^{\vec{\eta}}_{\vec{I},\vec{I}}$. This is shown for $\eta_{2i} = -\eta_{2i-1} =1$, for which the overlap is graphically represented,
\def\y{4.7} 
\def\wid{0.56}
\def\xinit{\wid*2+0.2}
\def\marg{0.1}
\begin{align}
& \frac{1}{d_\psi^{2n+1}} \begin{tikzpicture}[baseline=(current  bounding  box.center),
declare function =  {
	slope(\xi,\yi,\xj,\yj) = (\yi-\yj)/(\xi-\xj);
	yintercept(\xi,\yi,\xj,\yj) = \yi - \xi* slope(\xi,\yi,\xj,\yj);
	interx(\xi,\yi,\xj,\yj,\xk,\yk,\xl,\yl) = -(yintercept(\xi,\yi,\xj,\yj)-yintercept(\xk,\yk,\xl,\yl))/(slope(\xi,\yi,\xj,\yj)-slope(\xk,\yk,\xl,\yl));
	intery(\xi,\yi,\xj,\yj,\xk,\yk,\xl,\yl) =interx(\xi,\yi,\xj,\yj,\xk,\yk,\xl,\yl)* slope(\xi,\yi,\xj,\yj) +yintercept(\xi,\yi,\xj,\yj);
}]
\draw[gray, dashed,thick](0.0,0.3)--(\wid*13,0.3);
\node at (\xinit - \wid*2,0.7) {$\cdots$};
\foreach \pos in {\xinit} {
	\draw[gray,dashed, thick](\pos,0.3)--(\pos,0.6);
	\draw[-to,black,thick](\pos-\wid,0.5 + \wid*5/4)..controls(\pos-\wid,0.5) and(\pos+\wid,0.5)..(\pos+\wid,0.5 + \wid*5/4);
	\node at (\pos-\wid,0.5 + \wid*1.6) {$2i-1$};
	\node at (\pos+\wid,0.5 + \wid*1.6) {$2i-1'$};
}
\foreach \pos in {\xinit + \wid*4} {
	\draw[gray,dashed, thick](\pos,0.3)--(\pos,0.6);
	\draw[-to,black,thick](\pos-\wid,0.5 + \wid*5/4)..controls(\pos-\wid,0.5) and(\pos+\wid,0.5)..(\pos+\wid,0.5 + \wid*5/4);
	\node at (\pos-\wid,0.5 + \wid*1.6) {$\psi$};
	\node at (\pos+\wid,0.5 + \wid*1.6) {$\psi^\dagger$};
}
\foreach \pos in {\xinit + \wid*8} {
	\draw[gray,dashed, thick](\pos,0.3)--(\pos,0.6);
	\draw[-to,black,thick](\pos-\wid,0.5 + \wid*5/4)..controls(\pos-\wid,0.5) and(\pos+\wid,0.5)..(\pos+\wid,0.5 + \wid*5/4);
	\node at (\pos-\wid,0.5 + \wid*1.6) {$2i'$};
	\node at (\pos+\wid,0.5 + \wid*1.6) {$2i$};
}
\node at (\xinit + \wid*10,0.7) {$\cdots$};
\node at (\xinit + \wid*6,0.7) {$\cdots$};
\node at (\xinit + \wid*2,0.7) {$\cdots$};
\draw[gray, dashed,thick](0.0,\y-0.3)--(\wid*13,\y-0.3);
\node at (\xinit - \wid*2,\y-0.7) {$\cdots$};
\foreach \pos in {\xinit} {
	\draw[gray,dashed, thick](\pos,\y-0.3)--(\pos,\y-0.6);
	\draw[to-,black,thick](\pos-\wid,\y-0.5 - \wid*5/4)..controls(\pos-\wid,\y-0.5) and(\pos+\wid,\y-0.5)..(\pos+\wid,\y-0.5 - \wid*5/4);
	\node at (\pos-\wid,\y-0.5 - \wid*1.6) {$2i-1$};
	\node at (\pos+\wid,\y-0.5 - \wid*1.6) {$2i$};
}
\foreach \pos in {\xinit+\wid*4} {
	\draw[gray,dashed, thick](\pos,\y-0.3)--(\pos,\y-0.6);
	\draw[-to,black,thick](\pos-\wid,\y-0.5 - \wid*5/4)..controls(\pos-\wid,\y-0.5) and(\pos+\wid,\y-0.5)..(\pos+\wid,\y-0.5 - \wid*5/4);
	\node at (\pos-\wid,\y-0.5 - \wid*1.6) {$2i-1'$};
	\node at (\pos+\wid,\y-0.5 - \wid*1.6) {$2i'$};
}
\foreach \pos in {\xinit+\wid*8} {
	\draw[gray,dashed, thick](\pos,\y-0.3)--(\pos,\y-0.6);
	\draw[to-,black,thick](\pos-\wid,\y-0.5 - \wid*5/4)..controls(\pos-\wid,\y-0.5) and(\pos+\wid,\y-0.5)..(\pos+\wid,\y-0.5 - \wid*5/4);
	\node at (\pos-\wid,\y-0.5 - \wid*1.6) {$\psi$};
	\node at (\pos+\wid,\y-0.5 - \wid*1.6) {$\psi^\dagger$};
}
\node at (\xinit + \wid*10,\y-0.7) {$\cdots$};
\node at (\xinit + \wid*6,\y-0.7) {$\cdots$};
\draw[to-,black,thick](\xinit-\wid,0.5 + \wid*2.0)--(\xinit-\wid,\y-0.5 - \wid*2.0);
\draw[-to,black,thick](\xinit+\wid*5,0.5 + \wid*2.0)--(\xinit+\wid*9,\y-0.5 - \wid*2.0);
\def\Xi{\xinit+\wid*1}; \def\Yi{0.5 + \wid*2.0};\def\Xj{\xinit+\wid*3}; \def\Yj{\y-0.5 - \wid*2.0};
\def\Xk{\xinit+\wid*9}; \def\Yk{0.5 + \wid*2.0};\def\Xl{\xinit+\wid}; \def\Yl{\y-0.5 - \wid*2.0};
\draw[black,thick](\Xi,\Yi)--({interx(\Xi,\Yi,\Xj,\Yj,\Xk,\Yk,\Xl,\Yl)-\marg },{intery(\Xi,\Yi,\Xj,\Yj,\Xk,\Yk,\Xl,\Yl) -\marg*slope(\Xi,\Yi,\Xj,\Yj) });
\draw[-to,black,thick]({interx(\Xi,\Yi,\Xj,\Yj,\Xk,\Yk,\Xl,\Yl)+\marg },{intery(\Xi,\Yi,\Xj,\Yj,\Xk,\Yk,\Xl,\Yl) +\marg*slope(\Xi,\Yi,\Xj,\Yj)})--(\Xj,\Yj);
\def\Ytemp{\wid*\y*0.89};
\def\Xi{\xinit+\wid*7}; \def\Yi{0.5 + \wid*2.0};\def\Xj{\xinit+\wid*9}; \def\Yj{\Ytemp};
\def\Xk{\xinit+\wid*9}; \def\Yk{0.5 + \wid*2.0};\def\Xl{\xinit+\wid}; \def\Yl{\y-0.5 - \wid*2.0};
\draw[to-,black,thick](\Xi,\Yi)--({interx(\Xi,\Yi,\Xj,\Yj,\Xk,\Yk,\Xl,\Yl)-\marg },{intery(\Xi,\Yi,\Xj,\Yj,\Xk,\Yk,\Xl,\Yl) -\marg*slope(\Xi,\Yi,\Xj,\Yj) });
\draw[black,thick]({interx(\Xi,\Yi,\Xj,\Yj,\Xk,\Yk,\Xl,\Yl)+\marg },{intery(\Xi,\Yi,\Xj,\Yj,\Xk,\Yk,\Xl,\Yl) +\marg*slope(\Xi,\Yi,\Xj,\Yj)})--(\Xj,\Yj);
\def\Xi{\xinit+\wid*9}; \def\Yi{\Ytemp};\def\Xj{\xinit+\wid*5}; \def\Yj{\y-0.5 - \wid*2.0};
\def\Xk{\xinit+\wid*5}; \def\Yk{0.5 + \wid*2.0};\def\Xl{\xinit+\wid*9}; \def\Yl{\y-0.5 - \wid*2.0};
\draw[black,thick](\Xi,\Yi)--({interx(\Xi,\Yi,\Xj,\Yj,\Xk,\Yk,\Xl,\Yl)+\marg },{intery(\Xi,\Yi,\Xj,\Yj,\Xk,\Yk,\Xl,\Yl) +\marg*slope(\Xi,\Yi,\Xj,\Yj)});
\def\Xkk{\xinit+\wid*5}; \def\Ykk{0.5 + \wid*2.0};\def\Xll{\xinit+\wid*9}; \def\Yll{\y-0.5 - \wid*2.0};
\def\Xk{\xinit+\wid*3}; \def\Yk{0.5 + \wid*2.0};\def\Xl{\xinit+\wid*7}; \def\Yl{\y-0.5 - \wid*2.0};
\draw[black,thick]({interx(\Xi,\Yi,\Xj,\Yj,\Xk,\Yk,\Xl,\Yl)-\marg },{intery(\Xi,\Yi,\Xj,\Yj,\Xk,\Yk,\Xl,\Yl) -\marg*slope(\Xi,\Yi,\Xj,\Yj)})--(\Xj,\Yj);
\draw[black,thick]({interx(\Xi,\Yi,\Xj,\Yj,\Xkk,\Ykk,\Xll,\Yll)-\marg },{intery(\Xi,\Yi,\Xj,\Yj,\Xkk,\Ykk,\Xll,\Yll) -\marg*slope(\Xi,\Yi,\Xj,\Yj)})--({interx(\Xi,\Yi,\Xj,\Yj,\Xk,\Yk,\Xl,\Yl)+\marg },{intery(\Xi,\Yi,\Xj,\Yj,\Xk,\Yk,\Xl,\Yl) +\marg*slope(\Xi,\Yi,\Xj,\Yj)});
\def\Xi{\xinit+\wid*9}; \def\Yi{0.5 + \wid*2.0};\def\Xj{\xinit+\wid}; \def\Yj{\y-0.5 - \wid*2.0};
\def\Xk{\xinit+\wid*5}; \def\Yk{0.5 + \wid*2.0};\def\Xl{\xinit+\wid*9}; \def\Yl{\y-0.5 - \wid*2.0};
\draw[black,thick](\Xi,\Yi)--({interx(\Xi,\Yi,\Xj,\Yj,\Xk,\Yk,\Xl,\Yl)+\marg },{intery(\Xi,\Yi,\Xj,\Yj,\Xk,\Yk,\Xl,\Yl) +\marg*slope(\Xi,\Yi,\Xj,\Yj) });
\draw[-to,black,thick]({interx(\Xi,\Yi,\Xj,\Yj,\Xk,\Yk,\Xl,\Yl)-\marg },{intery(\Xi,\Yi,\Xj,\Yj,\Xk,\Yk,\Xl,\Yl) -\marg*slope(\Xi,\Yi,\Xj,\Yj)})--(\Xj,\Yj);
\def\Xi{\xinit+\wid*3}; \def\Yi{0.5 + \wid*2.0};\def\Xj{\xinit+\wid*7}; \def\Yj{\y-0.5 - \wid*2.0};
\def\Xk{\xinit+\wid*9}; \def\Yk{0.5 + \wid*2.0};\def\Xl{\xinit+\wid}; \def\Yl{\y-0.5 - \wid*2.0};
\draw[to-,black,thick](\Xi,\Yi)--({interx(\Xi,\Yi,\Xj,\Yj,\Xk,\Yk,\Xl,\Yl)-\marg },{intery(\Xi,\Yi,\Xj,\Yj,\Xk,\Yk,\Xl,\Yl) -\marg*slope(\Xi,\Yi,\Xj,\Yj) });
\draw[black,thick]({interx(\Xi,\Yi,\Xj,\Yj,\Xk,\Yk,\Xl,\Yl)+\marg },{intery(\Xi,\Yi,\Xj,\Yj,\Xk,\Yk,\Xl,\Yl) +\marg*slope(\Xi,\Yi,\Xj,\Yj)})--(\Xj,\Yj);
\end{tikzpicture}
\label{overlap-2}
\end{align}
Here $[\psi(x,t) \psi^\dagger(0,0)]_I$ is sandwiched by $\mathcal{A}^\dagger(t^{\pm}_{2i-1})$ and $\mathcal{A}(t^{\mp}_{2i})$ in the Keldysh ordering of $\mathcal{F}^{\vec{\eta}}$. In computing the overlap, $\mathcal{A}(t_{2i})$ has to move to the left of $\psi$ and $\psi^\dagger$ as in the operator ordering of $\mathcal{G}^{\vec{\eta}}_{\vec{I},\vec{I}}$. This is done with crossing of connections in Eq.~\eqref{overlap-2}.
When $0 <vt_{2i-1} +d \simeq vt_{2i} +d < vt-x$, the crossings are drawn as follows.
Connection $\psi^\dagger$ is on top of all the other connections, connections $(2i-1)$ and $(2i)$ are on top of all the other connections except $\psi^\dagger$, connection $\psi$ is on top of $(2i-1)'$ and  $(2i)'$ and below $\psi^\dagger$, $(2i-1)$ and $(2i)$; connections $(2i-1)'$, $(2i)'$, $\psi$, $(2i-1)$, $(2i)$, and $\psi^\dagger$ are placed in order from bottom to top. 
Hence the two loops, one composed of the connections $\psi$ and $\psi^\dagger$ for the main anyon fields of the correlator,
and the other of the connections $(2i-1)'$, $(2i)'$, $(2i-1)$, $(2i)$ for an anyon generated by the tunneling at QPC$_\textrm{A}$,
are linked in a non-contractible way. Untying the linked loops in Eq.~\eqref{overlap-2} into the two unlinked ones in Eq.~\eqref{overlap-1} indicates that the overlap is proportional to the monodromy $M$ of the anyons.
All the other cases of $\eta_{2i - 1} = - \eta_{2i}$ for some $i$, the overlap is evaluated similarily. Collecting the results for all the Keldysh indices, 
\begin{equation}
\begin{split}
\mathcal{F}^{\vec{\eta}} \simeq & \frac{1}{d_\psi^n}\prod_i a_{\eta_{2i-1},\eta_{2i}}(vt_{2i}+d) \mathcal{G}^{\vec{\eta}}_{\vec{I},\vec{I}} , 		\quad \quad
a_{\eta,\eta'}(v\bar{t})  =   \begin{cases}
M,& \text{if   } 0< v\bar{t}<vt-x \text{ } \& \text{ }\eta = -\eta' = -1 \\
M^*,& \text{if   } 0<v\bar{t}<vt-x \text{ } \& \text{ }\eta = -\eta' = 1 \\
1 &  \text{otherwise}.
\end{cases}
\end{split}
\end{equation}
Plugging this into Eq.~\eqref{introF} and taking summation over all the $\eta_i$'s and integration over $\frac{t_{2i-1}+t_{2i}}{2}$'s, we get 
\begin{equation}
\begin{split}
\mathcal{C}_n(x,t) \simeq &	\abs{\gamma_\text{A}}^{2n}\frac{(M-1)^n}{d_\psi^n n!} (t-\frac{x}{v})^n\langle[\psi(x,t)\psi^\dagger(0,0)]_I\rangle_\textrm{eq} \prod_i\int d(t_{2i-1} - t_{2i}) \frac{e^{i\omega (t_{2i-1} - t_{2i})}}{[\epsilon+ i(t_{2i-1} - t_{2i}) ]^{4h_\psi}}	\\
= &  \frac{1}{n!} [(M-1)(t-\frac{x}{v}) \frac{I_\text{inj}}{e^*}]^n \langle[\psi(x,t)\psi^\dagger(0,0)]_I\rangle_\textrm{eq}. \nonumber
\end{split}
\end{equation}
Doing the resummation $\expval{[\psi(x,t)\psi^\dagger(0,0)]_I}_\text{neq} = \sum_{n=0}^\infty \mathcal{C}_{n}(x,t)$, we obtain Eq.~\eqref{neqcorr}. A physical process contributing to $\mathcal{C}_1(x,t)$ is shown in Fig.~\ref{fig-corr} as an example. Equation~\eqref{neqcorr} is valid at long time $\abs{t-x/v} \gg \hbar/e^*V_\text{inj}$. And the perturbative expansion $\sum_{n=0}^\infty \mathcal{C}_{n}(x,t)$ converges in the weak-backscattering regime of the QPC. %

We emphasize that in Eq.~\eqref{neqcorr},
 the monodromy $M$ appears together with the trivial term $1$.
The combined factor $M-1$ originates from the fact that in every perturbation order, partial cancellation of the terms happens between different Keldysh indices in the domain of $0<v t_{2i-1}+d  \simeq v t_{2i}+d < vt-x$. The combined factor or the partial cancellation is understood, considering the collider of free electrons in the integer quantum Hall regime.
For free electrons, braiding is a trivial process, hence, Eq.~\eqref{overlap-2} is described by a disconnected Feynmann diagram (a vacuum bubble diagram) having $M=1$. This disconnected diagram does not contribute to observables for free electrons, since a partner disconnected diagram (a partner vacuum bubble) having the trivial factor $1$ pairwise appears and fully cancels the diagram, $M-1 = 0$, according to the linked cluster theorem of the many-body perturbation theory. By contrast, for anyons, the two diagrams, one giving the nontrivial monodromy $M \ne 1$ and its partner giving the trivial factor $1$, pairwise appear and cancel each other only partially, resulting in the combined factor $M-1 \ne 0$. 
Anyon processes involving the partial cancellation are called topological vacuum bubbles in Refs.~\cite{Han16-S,Blee19-S}.

\section{Subleading contributions to the correlator and currents}

We discuss and estimate certain subleading terms in Eqs.~\eqref{neqtunneling} and  \eqref{generalI}  [also in Eq.~\eqref{neqcorr}],
a subleading term corresponding to the coventional collision process of free electrons,
and another subleading term that regularizes the diverging behavior of $P_-$ in the cases of the non-Abelian anyons having $\Im[M]=0$. 

\subsection{Direct collision}
We consider the direct collision process, in which the particles injected at QPC$_\text{A/B}$ by the voltage $V_\textrm{A/B,inj}$ directly tunnel at QPC$_\text{C}$. 
For estimation of its contribution to the nonequilibrium correlator in Eq.~\eqref{neqtunneling}, it is sufficient to consider the lowest-order perturbation in the tunneling strength at QPC$_\text{A/B}$. 

This process has the configuration of time and Keldysh indices of $t_1 +d/v \simeq 0$, $t_2 +d/v \simeq t-x/v$, and $\eta_1 = -\eta_2 = -1$.
In the fusion basis $\mathcal{G}^{\vec{\eta}}_{\vec{I},\vec{I}}$ for this configuration, $\psi(-d,t_1)$ and $\psi^\dagger(0,0)$ fuse to the vacuum channel, and $\psi(x,t)$ and $\psi^\dagger(-d,t_2)$  also fuse to the vacuum channel.
The overlap of $\mathcal{F}^{\vec{\eta}}$ and  $\mathcal{G}^{\vec{\eta}}_{\vec{I},\vec{I}}$ for the lowest-order perturbation is shown~as 
\def\y{4.3} 
\def\wid{0.56}
\def\xinit{\wid*2+0.2}
\def\marg{0.1}
\begin{align}
& \frac{1}{d_\psi^{3}} \begin{tikzpicture}[baseline=(current  bounding  box.center),
declare function =  {
	slope(\xi,\yi,\xj,\yj) = (\yi-\yj)/(\xi-\xj);
	yintercept(\xi,\yi,\xj,\yj) = \yi - \xi* slope(\xi,\yi,\xj,\yj);
	interx(\xi,\yi,\xj,\yj,\xk,\yk,\xl,\yl) = -(yintercept(\xi,\yi,\xj,\yj)-yintercept(\xk,\yk,\xl,\yl))/(slope(\xi,\yi,\xj,\yj)-slope(\xk,\yk,\xl,\yl));
	intery(\xi,\yi,\xj,\yj,\xk,\yk,\xl,\yl) =interx(\xi,\yi,\xj,\yj,\xk,\yk,\xl,\yl)* slope(\xi,\yi,\xj,\yj) +yintercept(\xi,\yi,\xj,\yj);
}]
\draw[gray, dashed,thick](0.0,0.3)--(\wid*13,0.3);
\node at (\xinit - \wid*2,0.7) {$\cdots$};
\foreach \pos in {\xinit} {
	\draw[gray,dashed, thick](\pos,0.3)--(\pos,0.6);
	\draw[-to,black,thick](\pos-\wid,0.5 + \wid*5/4)..controls(\pos-\wid,0.5) and(\pos+\wid,0.5)..(\pos+\wid,0.5 + \wid*5/4);
	\node at (\pos-\wid,0.5 + \wid*1.6) {$1$};
	\node at (\pos+\wid,0.5 + \wid*1.6) {$1'$};
}
\foreach \pos in {\xinit + \wid*4} {
	\draw[gray,dashed, thick](\pos,0.3)--(\pos,0.6);
	\draw[-to,black,thick](\pos-\wid,0.5 + \wid*5/4)..controls(\pos-\wid,0.5) and(\pos+\wid,0.5)..(\pos+\wid,0.5 + \wid*5/4);
	\node at (\pos-\wid,0.5 + \wid*1.6) {$\psi$};
	\node at (\pos+\wid,0.5 + \wid*1.6) {$\psi^\dagger$};
}
\foreach \pos in {\xinit + \wid*8} {
	\draw[gray,dashed, thick](\pos,0.3)--(\pos,0.6);
	\draw[-to,black,thick](\pos-\wid,0.5 + \wid*5/4)..controls(\pos-\wid,0.5) and(\pos+\wid,0.5)..(\pos+\wid,0.5 + \wid*5/4);
	\node at (\pos-\wid,0.5 + \wid*1.6) {$2'$};
	\node at (\pos+\wid,0.5 + \wid*1.6) {$2$};
}
\draw[gray, dashed,thick](0.0,\y-0.3)--(\wid*13,\y-0.3);
\foreach \pos in {\xinit} {
	\draw[gray,dashed, thick](\pos,\y-0.3)--(\pos,\y-0.6);
	\draw[-to,black,thick](\pos-\wid,\y-0.5 - \wid*5/4)..controls(\pos-\wid,\y-0.5) and(\pos+\wid,\y-0.5)..(\pos+\wid,\y-0.5 - \wid*5/4);
	\node at (\pos-\wid,\y-0.5 - \wid*1.6) {$1'$};
	\node at (\pos+\wid,\y-0.5 - \wid*1.6) {$2'$};
}
\foreach \pos in {\xinit+\wid*4} {
	\draw[gray,dashed, thick](\pos,\y-0.3)--(\pos,\y-0.6);
	\draw[to-,black,thick](\pos-\wid,\y-0.5 - \wid*5/4)..controls(\pos-\wid,\y-0.5) and(\pos+\wid,\y-0.5)..(\pos+\wid,\y-0.5 - \wid*5/4);
	\node at (\pos-\wid,\y-0.5 - \wid*1.6) {$1$};
	\node at (\pos+\wid,\y-0.5 - \wid*1.6) {$\psi^\dagger$};
}
\foreach \pos in {\xinit+\wid*8} {
	\draw[gray,dashed, thick](\pos,\y-0.3)--(\pos,\y-0.6);
	\draw[to-,black,thick](\pos-\wid,\y-0.5 - \wid*5/4)..controls(\pos-\wid,\y-0.5) and(\pos+\wid,\y-0.5)..(\pos+\wid,\y-0.5 - \wid*5/4);
	\node at (\pos-\wid,\y-0.5 - \wid*1.6) {$\psi$};
	\node at (\pos+\wid,\y-0.5 - \wid*1.6) {$2$};
}
\draw[to-,black,thick](\xinit-\wid,0.5 + \wid*2.0)--(\xinit+3*\wid,\y-0.5 - \wid*2.0);
\draw[-to,black,thick](\xinit+9*\wid,0.5 + \wid*2.0)--(\xinit+9*\wid,\y-0.5 - \wid*2.0);
\draw[to-,black,thick](\xinit+3*\wid,0.5 + \wid*2.0)--(\xinit+7*\wid,\y-0.5 - \wid*2.0);
\def\Ymid{\y-\wid*4.9};
\draw[black,thick](\xinit+5*\wid,0.5 + \wid*2.0)--(\xinit+5*\wid,\Ymid-\marg);
\draw[-to,black,thick](\xinit+5*\wid,\Ymid+\marg)--(\xinit+5*\wid,\y-0.5 - \wid*2.0);
\def\Xi{\xinit+\wid*1}; \def\Yi{0.5 + \wid*2.0};\def\Xj{\xinit-\wid}; \def\Yj{\y-0.5 - \wid*2.0};
\def\Xk{\xinit-\wid}; \def\Yk{0.5 + \wid*2.0};\def\Xl{\xinit+3*\wid}; \def\Yl{\y-0.5 - \wid*2.0};
\draw[black,thick](\Xi,\Yi)--({interx(\Xi,\Yi,\Xj,\Yj,\Xk,\Yk,\Xl,\Yl)+\marg },{intery(\Xi,\Yi,\Xj,\Yj,\Xk,\Yk,\Xl,\Yl) +\marg*slope(\Xi,\Yi,\Xj,\Yj) });
\draw[-to,black,thick]({interx(\Xi,\Yi,\Xj,\Yj,\Xk,\Yk,\Xl,\Yl)-\marg },{intery(\Xi,\Yi,\Xj,\Yj,\Xk,\Yk,\Xl,\Yl) -\marg*slope(\Xi,\Yi,\Xj,\Yj)})--(\Xj,\Yj);
\def\Xi{\xinit+\wid*7}; \def\Yi{0.5 + \wid*2.0};\def\Xj{\xinit+\wid}; \def\Yj{\y-0.5 - \wid*2.0};
\def\Xk{\xinit+\wid*3}; \def\Yk{0.5 + \wid*2.0};\def\Xl{\xinit+7*\wid}; \def\Yl{\y-0.5 - \wid*2.0};
\draw[black,thick](\Xi,\Yi)--({interx(\Xi,\Yi,\Xj,\Yj,\Xk,\Yk,\Xl,\Yl)+3*\marg },{intery(\Xi,\Yi,\Xj,\Yj,\Xk,\Yk,\Xl,\Yl) +3*\marg*slope(\Xi,\Yi,\Xj,\Yj)});
\def\Xkk{\xinit-\wid}; \def\Ykk{0.5 + \wid*2.0};\def\Xll{\xinit+\wid*3}; \def\Yll{\y-0.5 - \wid*2.0};
\def\margk{15}
\draw[black,thick]({interx(\Xi,\Yi,\Xj,\Yj,\Xk,\Yk,\Xl,\Yl)-\margk*\marg },{intery(\Xi,\Yi,\Xj,\Yj,\Xk,\Yk,\Xl,\Yl) -\margk*\marg*slope(\Xi,\Yi,\Xj,\Yj)})--(\Xj,\Yj);
\def\margk{9}
\draw[black,thick]({interx(\Xi,\Yi,\Xj,\Yj,\Xk,\Yk,\Xl,\Yl)-\marg },{intery(\Xi,\Yi,\Xj,\Yj,\Xk,\Yk,\Xl,\Yl) -\marg*slope(\Xi,\Yi,\Xj,\Yj)})--(\Xj+\marg*\margk,{\Yj+\margk*\marg*slope(\Xi,\Yi,\Xj,\Yj)});
\end{tikzpicture}
=\frac{e^{-i2\pi h_\psi}}{d_\psi^2},
\label{overlap-conv}
\end{align}
where $e^{-i2\pi h_\psi}$ is the topological spin of the anyon.
The  process contributes to the nonequilibrium correlator as
\begin{equation}
\begin{split}
\frac{d^{-2}_\psi\abs{\gamma_\text{A}}^2e^{-i2\pi h_\psi}}{[\epsilon-i(t-x/v)]^{2h_\psi}}\int dt_1 \frac{e^{i\omega t_1}}{[\epsilon +i(t_1+d/v)]^{2h_\psi}}\int dt_2\frac{e^{-i\omega t_2}}{[\epsilon -i(t_2-t +(d+x)/v )]^{2h_\psi}} = \frac{\abs{\gamma_\text{A}}^2\omega^{4h_\psi-2}}{\Gamma(2h_\psi)^2d_\psi^2} \frac{e^{-i\omega (t-x/v)}}{[\epsilon+i(t-x/v)]^{2h_\psi}}.
\end{split}\label{corr-conv}
\end{equation}
We notice that the contribution smoothly recovers the free electron result of integer quantum Hall edges as $h_\psi \rightarrow 1/2$ and $d_\psi \rightarrow 1$.
This shows that this process corresponds to the conventional collision process of free electrons.
 
We further estimate the contribution of this process to $I_\textrm{T}$, $\langle \delta I_\textrm{T}^2 \rangle$, and $\langle \delta I_\textrm{A} \delta I_\textrm{B} \rangle$ as $ O\Big( V_\text{A/B,inj}^{4\delta -3} \Big)$.
We compare it with the leading contribution from the time-domain interference involving braiding,
\begin{equation}
\frac{\text{Contribution from the direct collision process}}{\text{Contribution from the leading term in Eq.~\eqref{generalI}} } =  	O\Big((\frac{\hbar}{e^{*2}}\frac{I_\text{inj}}{V_\text{inj}})^{2-2\delta}\Big).
\label{order-conv}
\end{equation}
Hence, for $\delta < 1$, the collision process is sub-dominant in comparison with the time-domain interference, as discussed in the main text. We remark that because of the exponential factor $e^{-i\omega (t-x/v)}$ in Eq.~\eqref{corr-conv}, the contribution of the direct collision process to $I_\textrm{T} = e^* \int_{-\infty}^\infty dt \expval{\comm{\mathcal{T}^\dagger(0)}{\mathcal{T}(t)}}_\text{neq}$ and $\expval{\delta I_\textrm{T}^2} = e^{*2}\int_{-\infty}^\infty dt \expval{\acomm{\mathcal{T}^\dagger(0)}{\mathcal{T}(t)}}_\text{neq}$ is mainly determined by the short time window of $t\simeq 0$. Braiding cannot happen in the direct process, since it requires other time windows of   $0 < t_{2i-1} + d/v\simeq t_{2i} + d/v < t$ or $t < t_{2i-1} + d/v \simeq t_{2i}+ d/v < 0$.

\subsection{Subleading braiding process}
\label{subsection_subleading}

We consider non-Abelian anyons having $\Im[M]=0$. 
In this particular case, the time-domain interference involving the braiding in Eq.~\eqref{overlap-2} provides the dominant contribution to Eq.~\eqref{neqtunneling}, but its contribution to $I_\textrm{T}$ in Eq.~\eqref{generalI} vanishes, leading to the divergence of $P_-$ in the main text.
The divergence is regularized by the process that gives a subleading contribution to Eq.~\eqref{neqtunneling} and
the biggest contribution to $I_\textrm{T}$ when $\Im[M]=0$.
We below discuss this subleading process.
 
\begin{figure}[b]
	\centering
	\includegraphics[width = 0.23 \textwidth]{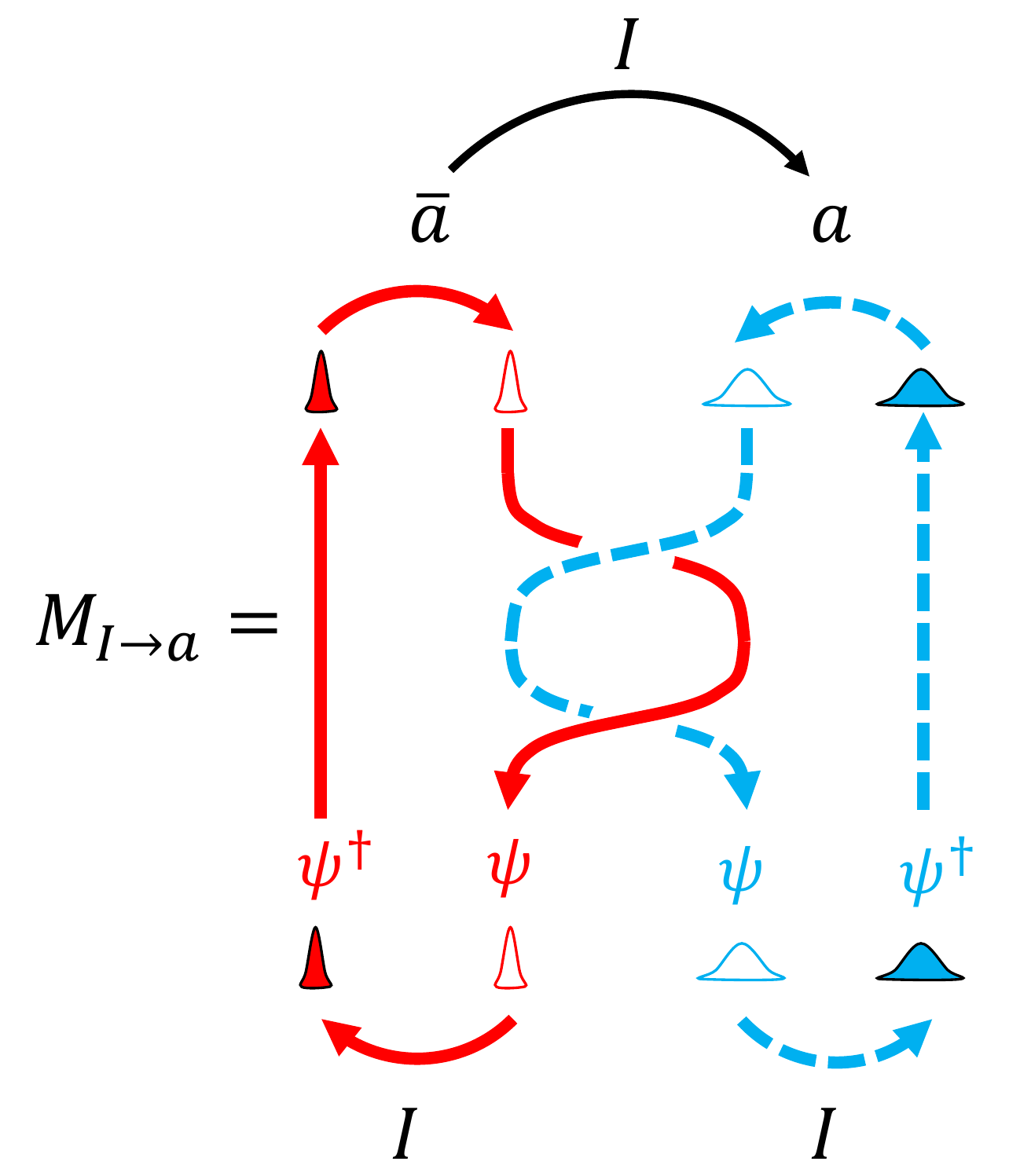}
	\caption{Monodromy $M_{I \rightarrow a}$. Two particle-hole pairs of $\psi$ anyons are initially split from the vacuum. After the braiding, they finally fuse into channels $\bar{a}$ and $a$. The transition amplitude is  $M_{I \rightarrow a}$.
The monodromy $M$ discussed in the main text is $M_{I \rightarrow I}$.
}\label{SFig-diffmono}
\end{figure}

There are other time-domain interference processes identical to the braiding process of Eq.~\eqref{overlap-2}, having the pairings of $t_{2i'-1}\simeq t_{2i'}$ in the time configuration, except that its fusion basis $\mathcal{G}^{\vec{\eta}}_{\vec{a},\vec{b}}$ differs from the basis $\mathcal{G}^{\vec{\eta}}_{\vec{I},\vec{I}}$ of Eq.~\eqref{overlap-2}.  
Among those processes, we focus on the one in which the main fields $\psi$ and $\psi^\dagger$ of the correlator is fused into a fusion state $a_{2n+1}=a \neq I$ different from the vacuum $I$ while the fusion state $a_{2i-1}$ of an injected anyon,  described by $\mathcal{A}^\dagger(t_{2i-1})$ and $\mathcal{A}(t_{2i})$ at $t_{2i-1}\simeq t_{2i}$, is in the fusion channel $a_{2i-1} =\bar{a}$ on the anti-anyon of $a$; all the other fusion states $a_{j \ne 2i-1, 2n+1}$ are the vacuum $I$, and hence, the intermediate fusion states $b_i$ are either $a$ ($b_{2i-1} = \cdots= b_{2n} = a$) or the vacuum ($b_1 = b_2 = \cdots = b_{2i-2} = I$).
The overlap of this fusion basis $\mathcal{G}^{\vec{\eta}}_{\vec{a},\vec{b}}$ and the conformal block $\mathcal{F}^{\vec{\eta}}$ is depicted,
\def\y{4.7} 
\def\wid{0.56}
\def\xinit{\wid*2+0.2}
\def\marg{0.1}
\begin{align}
& \frac{1}{d_\psi^{2n+1}} \begin{tikzpicture}[baseline=(current  bounding  box.center),
declare function =  {
	slope(\xi,\yi,\xj,\yj) = (\yi-\yj)/(\xi-\xj);
	yintercept(\xi,\yi,\xj,\yj) = \yi - \xi* slope(\xi,\yi,\xj,\yj);
	interx(\xi,\yi,\xj,\yj,\xk,\yk,\xl,\yl) = -(yintercept(\xi,\yi,\xj,\yj)-yintercept(\xk,\yk,\xl,\yl))/(slope(\xi,\yi,\xj,\yj)-slope(\xk,\yk,\xl,\yl));
	intery(\xi,\yi,\xj,\yj,\xk,\yk,\xl,\yl) =interx(\xi,\yi,\xj,\yj,\xk,\yk,\xl,\yl)* slope(\xi,\yi,\xj,\yj) +yintercept(\xi,\yi,\xj,\yj);
}]
\draw[gray, dashed,thick](0.0,0.3)--(\wid*13,0.3);
\node at (\xinit - \wid*2,0.7) {$\cdots$};
\foreach \pos in {\xinit} {
	\draw[gray,dashed, thick](\pos,0.3)--(\pos,0.6);
	\draw[-to,black,thick](\pos-\wid,0.5 + \wid*5/4)..controls(\pos-\wid,0.5) and(\pos+\wid,0.5)..(\pos+\wid,0.5 + \wid*5/4);
	\node at (\pos-\wid,0.5 + \wid*1.6) {$2i-1$};
	\node at (\pos+\wid,0.5 + \wid*1.6) {$2i-1'$};
}
\foreach \pos in {\xinit + \wid*4} {
	\draw[gray,dashed, thick](\pos,0.3)--(\pos,0.6);
	\draw[-to,black,thick](\pos-\wid,0.5 + \wid*5/4)..controls(\pos-\wid,0.5) and(\pos+\wid,0.5)..(\pos+\wid,0.5 + \wid*5/4);
	\node at (\pos-\wid,0.5 + \wid*1.6) {$\psi$};
	\node at (\pos+\wid,0.5 + \wid*1.6) {$\psi^\dagger$};
}
\foreach \pos in {\xinit + \wid*8} {
	\draw[gray,dashed, thick](\pos,0.3)--(\pos,0.6);
	\draw[-to,black,thick](\pos-\wid,0.5 + \wid*5/4)..controls(\pos-\wid,0.5) and(\pos+\wid,0.5)..(\pos+\wid,0.5 + \wid*5/4);
	\node at (\pos-\wid,0.5 + \wid*1.6) {$2i'$};
	\node at (\pos+\wid,0.5 + \wid*1.6) {$2i$};
}
\node at (\xinit + \wid*10,0.7) {$\cdots$};
\node at (\xinit + \wid*6,0.7) {$\cdots$};
\node at (\xinit + \wid*2,0.7) {$\cdots$};
\node at (\xinit+\wid*6 - \wid*2,\y-0.) {$a$};
\draw[gray, dashed,thick](0.0,\y-0.3)--(\wid*13,\y-0.3);
\draw[black,thick](\xinit,\y-0.3)--(\xinit+\wid*8,\y-0.3);
\node at (\xinit - \wid*2,\y-0.7) {$\cdots$};
\foreach \pos in {\xinit} {
	\draw[to-,black, thick](\pos,\y-0.3)--(\pos,\y-0.66);
	\draw[to-,black,thick](\pos-\wid,\y-0.5 - \wid*5/4)..controls(\pos-\wid,\y-0.5) and(\pos+\wid,\y-0.5)..(\pos+\wid,\y-0.5 - \wid*5/4);
	\node at (\pos-\wid,\y-0.5 - \wid*1.6) {$2i-1$};
	\node at (\pos+\wid,\y-0.5 - \wid*1.6) {$2i$};
}
\foreach \pos in {\xinit+\wid*4} {
	\draw[gray,dashed, thick](\pos,\y-0.3)--(\pos,\y-0.6);
	\draw[-to,black,thick](\pos-\wid,\y-0.5 - \wid*5/4)..controls(\pos-\wid,\y-0.5) and(\pos+\wid,\y-0.5)..(\pos+\wid,\y-0.5 - \wid*5/4);
	\node at (\pos-\wid,\y-0.5 - \wid*1.6) {$2i-1'$};
	\node at (\pos+\wid,\y-0.5 - \wid*1.6) {$2i'$};
}
\foreach \pos in {\xinit+\wid*8} {
	\draw[-to,black, thick](\pos,\y-0.3)--(\pos,\y-0.66);
	\draw[to-,black,thick](\pos-\wid,\y-0.5 - \wid*5/4)..controls(\pos-\wid,\y-0.5) and(\pos+\wid,\y-0.5)..(\pos+\wid,\y-0.5 - \wid*5/4);
	\node at (\pos-\wid,\y-0.5 - \wid*1.6) {$\psi$};
	\node at (\pos+\wid,\y-0.5 - \wid*1.6) {$\psi^\dagger$};
}
\node at (\xinit + \wid*10,\y-0.7) {$\cdots$};
\node at (\xinit + \wid*6,\y-0.7) {$\cdots$};
\draw[to-,black,thick](\xinit-\wid,0.5 + \wid*2.0)--(\xinit-\wid,\y-0.5 - \wid*2.0);
\draw[-to,black,thick](\xinit+\wid*5,0.5 + \wid*2.0)--(\xinit+\wid*9,\y-0.5 - \wid*2.0);
\def\Xi{\xinit+\wid*1}; \def\Yi{0.5 + \wid*2.0};\def\Xj{\xinit+\wid*3}; \def\Yj{\y-0.5 - \wid*2.0};
\def\Xk{\xinit+\wid*9}; \def\Yk{0.5 + \wid*2.0};\def\Xl{\xinit+\wid}; \def\Yl{\y-0.5 - \wid*2.0};
\draw[black,thick](\Xi,\Yi)--({interx(\Xi,\Yi,\Xj,\Yj,\Xk,\Yk,\Xl,\Yl)-\marg },{intery(\Xi,\Yi,\Xj,\Yj,\Xk,\Yk,\Xl,\Yl) -\marg*slope(\Xi,\Yi,\Xj,\Yj) });
\draw[-to,black,thick]({interx(\Xi,\Yi,\Xj,\Yj,\Xk,\Yk,\Xl,\Yl)+\marg },{intery(\Xi,\Yi,\Xj,\Yj,\Xk,\Yk,\Xl,\Yl) +\marg*slope(\Xi,\Yi,\Xj,\Yj)})--(\Xj,\Yj);
\def\Ytemp{\wid*\y*0.89};
\def\Xi{\xinit+\wid*7}; \def\Yi{0.5 + \wid*2.0};\def\Xj{\xinit+\wid*9}; \def\Yj{\Ytemp};
\def\Xk{\xinit+\wid*9}; \def\Yk{0.5 + \wid*2.0};\def\Xl{\xinit+\wid}; \def\Yl{\y-0.5 - \wid*2.0};
\draw[to-,black,thick](\Xi,\Yi)--({interx(\Xi,\Yi,\Xj,\Yj,\Xk,\Yk,\Xl,\Yl)-\marg },{intery(\Xi,\Yi,\Xj,\Yj,\Xk,\Yk,\Xl,\Yl) -\marg*slope(\Xi,\Yi,\Xj,\Yj) });
\draw[black,thick]({interx(\Xi,\Yi,\Xj,\Yj,\Xk,\Yk,\Xl,\Yl)+\marg },{intery(\Xi,\Yi,\Xj,\Yj,\Xk,\Yk,\Xl,\Yl) +\marg*slope(\Xi,\Yi,\Xj,\Yj)})--(\Xj,\Yj);
\def\Xi{\xinit+\wid*9}; \def\Yi{\Ytemp};\def\Xj{\xinit+\wid*5}; \def\Yj{\y-0.5 - \wid*2.0};
\def\Xk{\xinit+\wid*5}; \def\Yk{0.5 + \wid*2.0};\def\Xl{\xinit+\wid*9}; \def\Yl{\y-0.5 - \wid*2.0};
\draw[black,thick](\Xi,\Yi)--({interx(\Xi,\Yi,\Xj,\Yj,\Xk,\Yk,\Xl,\Yl)+\marg },{intery(\Xi,\Yi,\Xj,\Yj,\Xk,\Yk,\Xl,\Yl) +\marg*slope(\Xi,\Yi,\Xj,\Yj)});
\def\Xkk{\xinit+\wid*5}; \def\Ykk{0.5 + \wid*2.0};\def\Xll{\xinit+\wid*9}; \def\Yll{\y-0.5 - \wid*2.0};
\def\Xk{\xinit+\wid*3}; \def\Yk{0.5 + \wid*2.0};\def\Xl{\xinit+\wid*7}; \def\Yl{\y-0.5 - \wid*2.0};
\draw[black,thick]({interx(\Xi,\Yi,\Xj,\Yj,\Xk,\Yk,\Xl,\Yl)-\marg },{intery(\Xi,\Yi,\Xj,\Yj,\Xk,\Yk,\Xl,\Yl) -\marg*slope(\Xi,\Yi,\Xj,\Yj)})--(\Xj,\Yj);
\draw[black,thick]({interx(\Xi,\Yi,\Xj,\Yj,\Xkk,\Ykk,\Xll,\Yll)-\marg },{intery(\Xi,\Yi,\Xj,\Yj,\Xkk,\Ykk,\Xll,\Yll) -\marg*slope(\Xi,\Yi,\Xj,\Yj)})--({interx(\Xi,\Yi,\Xj,\Yj,\Xk,\Yk,\Xl,\Yl)+\marg },{intery(\Xi,\Yi,\Xj,\Yj,\Xk,\Yk,\Xl,\Yl) +\marg*slope(\Xi,\Yi,\Xj,\Yj)});
\def\Xi{\xinit+\wid*9}; \def\Yi{0.5 + \wid*2.0};\def\Xj{\xinit+\wid}; \def\Yj{\y-0.5 - \wid*2.0};
\def\Xk{\xinit+\wid*5}; \def\Yk{0.5 + \wid*2.0};\def\Xl{\xinit+\wid*9}; \def\Yl{\y-0.5 - \wid*2.0};
\draw[black,thick](\Xi,\Yi)--({interx(\Xi,\Yi,\Xj,\Yj,\Xk,\Yk,\Xl,\Yl)+\marg },{intery(\Xi,\Yi,\Xj,\Yj,\Xk,\Yk,\Xl,\Yl) +\marg*slope(\Xi,\Yi,\Xj,\Yj) });
\draw[-to,black,thick]({interx(\Xi,\Yi,\Xj,\Yj,\Xk,\Yk,\Xl,\Yl)-\marg },{intery(\Xi,\Yi,\Xj,\Yj,\Xk,\Yk,\Xl,\Yl) -\marg*slope(\Xi,\Yi,\Xj,\Yj)})--(\Xj,\Yj);
\def\Xi{\xinit+\wid*3}; \def\Yi{0.5 + \wid*2.0};\def\Xj{\xinit+\wid*7}; \def\Yj{\y-0.5 - \wid*2.0};
\def\Xk{\xinit+\wid*9}; \def\Yk{0.5 + \wid*2.0};\def\Xl{\xinit+\wid}; \def\Yl{\y-0.5 - \wid*2.0};
\draw[to-,black,thick](\Xi,\Yi)--({interx(\Xi,\Yi,\Xj,\Yj,\Xk,\Yk,\Xl,\Yl)-\marg },{intery(\Xi,\Yi,\Xj,\Yj,\Xk,\Yk,\Xl,\Yl) -\marg*slope(\Xi,\Yi,\Xj,\Yj) });
\draw[black,thick]({interx(\Xi,\Yi,\Xj,\Yj,\Xk,\Yk,\Xl,\Yl)+\marg },{intery(\Xi,\Yi,\Xj,\Yj,\Xk,\Yk,\Xl,\Yl) +\marg*slope(\Xi,\Yi,\Xj,\Yj)})--(\Xj,\Yj);
\end{tikzpicture}
\label{overlap-subleading}
\end{align}
showing that the two loops, one composed of  the connections $\psi$ and $\psi^\dagger$ for the main anyon fields of the correlator,
and the other of the connections $(2i-1)'$, $(2i)'$, $(2i-1)$, $(2i)$ for an anyon injected through QPC$_\textrm{A}$,
are linked in a non-contractible way, similarly to Eq.~\eqref{overlap-2}.
However, untying the linked loops in Eq.~\eqref{overlap-subleading} into two unlinked ones indicates that the overlap is proportional to another monodromy $M_{I\rightarrow a}$ of the anyons, which differs from the monodromy $M$ involved in Eq.~\eqref{overlap-2} and discussed in the main text. $M_{I\rightarrow a}$ is depicted in Fig.~\ref{SFig-diffmono}.

The value of $M_{I \rightarrow a}$ depends on non-Abelian anyons. 
For $\text{SU}(2)_k$ anyons of the anti-Read-Rezayi (ARR) state at the level $k$, the fusion rule for $j=1/2$ anyons is $\frac{1}{2} \times \frac{1}{2} = 0 + 1$, hence, there are two possible $a$'s, $j=0$ and $j=1$. The fusion channel $j=0$ is the vacuum field $I$, and the case of $M_{I \rightarrow a=I}$ is discussed in Eq.~\eqref{overlap-2}.
The case of the $j=1$ fusion channel has the monodromy  $M_{I \rightarrow j= 1}=\frac{(e^{-2\pi i/(2+k)}-1) [\Gamma(k/(2+k))]^2}{\Gamma((-1+k)/(2+k))\Gamma((1+k)/(2+k))}$. 
The anti-Paffian state at $\nu = 5/2$ (the ARR state of level $k=2$) has
$M_{I \rightarrow j= 1}= - e^{i \pi /4}$  and 
the ARR state of level $k=3$ at $\nu = 12/5$ has $M_{I \rightarrow j= 1}=\frac{(e^{-2\pi i/5}-1) [\Gamma(3/5)]^2}{\Gamma(2/5)\Gamma(4/5)}$.
On the other hand, for the Ising anyon in the particle-hole symmetric Pfaffian state, the fusion rule is $\sigma \times \sigma = I + \Psi $, hence, there are also two possible $a$'s, the vacuum field $I$ and the fermion channel $\Psi$. The case of the fermion channel $\Psi$ has the monodromy $M_{I \rightarrow \Psi}=e^{i\pi/4}$.  
For all these non-Abelian anyons, $\Im[M_{I \rightarrow a}] \neq 0$, so the time-domain interference process involving the braiding in Fig.~\ref{SFig-diffmono} provides non-vanishing contribution to $I_\textrm{T}$.

We compute the contribution from the process in Eq.~\eqref{overlap-subleading} to the $n$-th Keldysh perturbation expansion term $\mathcal{C}_{n}(x,t)$ of the non-equilibrium correlator $\expval{[\psi_\textrm{A}(x,t)\psi_\textrm{A}^\dagger(0,0)]_I}_\text{neq}$.
The contribution is expressed as 
 \begin{equation}
\begin{split}
\frac{(M-1)^{n-1}}{(n-1)!}[\frac{I_\text{inj}}{e^*}(t-\frac{x}{v})]^{n-1}\abs{\gamma_\text{A}}^2\frac{M_{I\rightarrow a}}{d_\psi}\int d(t_{2i-1} - t_{2i}) \frac{e^{i\omega (t_{2i-1} - t_{2i})}}{[\epsilon+ i(t_{2i-1} - t_{2i}) ]^{4h_\psi-h_a}}  \\ \times \frac{1}{[\epsilon+i(t-x/v)]^{2h_{\psi}-h_{a}}} \int_{-d/v}^{t-(x+d)/v} d\tilde t \frac{1}{[\epsilon+i(\tilde{t}-t+(x+d)/v)]^{h_a}}\frac{1}{[\epsilon+i(\tilde{t}+d/v)]^{h_a}},
\end{split}
\end{equation}	
where $\tilde{t} = (t_{2i-1}+t_{2i})/2$.
In the derivation of this expression, we have multiplied the factor $n$, since there are $n$ possibilities of choosing $a_{2i-1}$ among $a_1$, $a_3$, $\cdots$, $a_{2n-1}$,
integrated out all the time indices $t_j$'s except the times $t_{2i-1}$ and $t_{2i}$ of the anyons fused into the channel $a$, applied OPE at $t_{2i-1}\simeq t_{2i}$ for connections $(2i-1)$ and $(2i)$, and then used the three-point function of the primary fields~\cite{Francesco-S}. Performing the integration over $t_{2i-1}-t_{2i}$ and $\tilde{t}$, we obtain
\begin{equation}
\begin{split}
\frac{(M-1)^{n-1}}{(n-1)!} [\frac{I_\text{inj}}{e^*}(t-\frac{x}{v})]^{n-1}\frac{M_{I\rightarrow a}}{d_\psi}\frac{2\pi\abs{\gamma_A}^2\omega^{4h_\psi -h_a -1}}{\Gamma(4h_\psi -h_a)} \frac{\Gamma(1-h_a)^2}{\Gamma(2-2h_a)}(t-\frac{x}{v})^{1-2h_a}\frac{1}{[\epsilon+i(t-x/v)]^{2h_{\psi}-h_{a}}}.		 
\end{split}
\end{equation}
Collecting this contribution to $\mathcal{C}_{n}(x,t)$ for different $n$'s
and doing the resummation of the contributions over $n$, we obtain the contribution from the process in Eq.~\eqref{overlap-subleading} and Fig.~\ref{SFig-diffmono} to the nonequilibrium correlator in Eq.~\eqref{neqcorr},
\begin{equation} \label{biggestsubleading}
\begin{split}
M_{I\rightarrow a} e^{(M-1)\frac{I_\text{inj}}{e^*}(t-x/v)}\frac{\Gamma(1-h_a)^2\Gamma(4h_\psi)}{\Gamma(2-2h_a)\Gamma(4h_\psi-h_a)}\frac{I_\text{inj}}{e^*} (e^*V_\text{inj}/\hbar)^{-{h_a}}  \frac{(t-x/v)^{1-2h_a}}{[\epsilon+i(t-x/v)]^{2h_{\psi}-h_{a}}}
\end{split}
\end{equation}
at $t - x > 0$.
This subleading contribution results in a term of $O\Big(I_\text{A/B,inj}^{2\delta+h_a-1}V_\text{A/B,inj}^{-h_a}\Big)$ in $I_\text{T} $ and $\expval{\delta I_\text{T}^2}$.
It is compared with the contribution from the leading term of Eq.~\eqref{neqcorr} originating from the process in Eq.~\eqref{overlap-2},
\begin{equation}
\frac{\text{Contribution from the process of } \mathcal{G}^{\vec{\eta}}_{\vec{a},\vec{b}}|_{a_{2n+1} = a, a_{2i-1} = \bar{a}} \text{ to } \expval{\delta I_\text{T}^2}}{\text{Contribution from the leading term of Eq.~\eqref{generalI} to } \expval{\delta I_\text{T}^2}} =O\Big((\frac{\hbar}{e^{*2}}\frac{I_\text{A/B,inj}}{V_\text{A/B,inj}})^{h_a}\Big).	\label{order-fusa}
\end{equation}
As $h_a >0$, the process involving the monodromy $M_{I \to a}$ indeed gives smaller contribution to $ \expval{\delta I_\text{T}^2}$ than the leading term of Eq.~\eqref{neqcorr}. The value of $h_a$ is $h_{j=1} = 2/(k+2)$ for $\text{SU}(2)_k$ anyons and $h_\Psi = 1/2$ for Ising anyons. 

Similarly, for anyons having $\textrm{Im} [M] \ne 0$, the process involving $M_{I \to a}$ negligibly contributes to $I_\text{T}$ in comparison with the leading term of Eq.~\eqref{neqcorr}, as the ratio of their contributions is $O\Big((\frac{\hbar}{e^{*2}}\frac{I_\text{A/B,inj}}{V_\text{A/B,inj}})^{h_a}\Big)$ as in Eq.~\eqref{order-fusa}.
For non-Abelian anyons having $\textrm{Im} [M] = 0$, however, this process gives the leading contribution to $I_\text{T}$
(bigger than the contribution from the direct collision in Eq.~\eqref{corr-conv}), therefore, $e^* I_\textrm{T} / \expval{\delta I_\textrm{T}^2} \sim O\Big((\frac{\hbar}{e^{*2}}\frac{I_\text{A/B,inj}}{V_\text{A/B,inj}})^{h_a}\Big)$, resulting in the divergence of $P_-$.

\section{Derivation of cross correlation $\expval{\delta I_\textrm{A} \delta I_\textrm{B}}$}\label{sec-crossnoise}
We derive the cross correlation $\expval{\delta I_\text{A} \delta I_\text{B}}$ at zero temperature. 
Charge conservation causes the relation $I_{\alpha =\textrm{A/B}} = I_{\alpha,\text{inj}}  - I_\textrm{T}$ between different currents.
The cross correlation is decomposed,
\begin{equation}
\begin{split}
\expval{\delta I_\text{A} \delta I_\text{B}}= & - \expval{\delta I_\textrm{T}^2 } + \expval{\delta I_{\text{A,inj}} \delta I_\textrm{T}} - 	\expval{\delta I_{\text{B,inj}} \delta I_\textrm{T}}   + \expval{\delta I_{\text{A,inj}} \delta I_{\text{B,inj}} }.
\end{split}
\end{equation}
$\expval{\delta I_{\text{A,inj}} \delta I_{\text{B,inj}}}$ is nonzero when anyons are fractionalized into upstream and downstream parts upon tunneling at QPC$_\textrm{C}$, but it is negligibly small when QPC$_\textrm{C}$ is in the weak tunneling regime (see Sec. \ref{sec-back}). 
 
We prove that for the case of $\textrm{Im} [M] \ne 0$, the zero-frequency correlation $\expval{\delta I_{\alpha,\text{inj}}\delta I_\textrm{T}}$ between the injection current at QPC$_{\alpha = \textrm{A}, \textrm{B}}$ and the tunneling current at QPC$_\textrm{C}$ satisfies
\begin{equation}\label{AppA-result}
\begin{split}
\expval{\delta I_{\alpha,\text{inj}} \delta I_\textrm{T}} = e^* I_{\alpha, \text{inj}}\frac{\partial I_\textrm{T}}{\partial I_{\alpha,\text{inj}}}.
\end{split}
\end{equation}
The case of $\textrm{Im}[M] = 0$ will be discussed separately.
We apply the perturbative expansion, over arbitrary orders of the tunneling strength at QPC$_\textrm{A}$ and the lowest order of the tunneling strength at QPC$_\textrm{C}$, to the correlation,
\begin{equation}
\begin{split}
\expval{\delta I_{\text{A,inj}} \delta I_\textrm{T}} = & \frac{1}{2}\int dt_1\expval{\acomm{\delta I_{ \text{A,inj}} (t_1)}{\delta I_\textrm{T}(0)}} 	= \frac{1}{4} \int dt_1 \sum_{\eta_0,\eta_1=\pm}\expval{T_K\{\delta I_\textrm{T}(0^{\eta_0}) \delta I_{\text{A,inj}}(t^{\eta_1}) \}}  \equiv \sum_n \expval{\delta I_{\text{A,inj}} \delta I_\textrm{T}}_n.
\end{split}
\end{equation}
In the second equality, we expressed the anti-commutator using the Keldysh ordering. 
$\expval{\delta I_{ \text{A,inj}} \delta I_\textrm{T}}_n$ represents the term of the order of $\abs{\gamma_\text{A}}^{2n} \abs{\gamma_\text{C}}^2$.  Employing the Keldysh method, we obtain
\begin{widetext}
\begin{equation}\label{cross-derv}
\begin{split}
& \expval{ \delta I_{\text{A,inj}} \delta I_\textrm{T}}_n 
= \frac{c_n}{4} \sum_{\eta_0,\eta,\eta_j} \eta \int dt dt_1 \cdots dt_{2n} \langle T_K\{I_\textrm{T}(0^{\eta_0})H_\textrm{T}(t^\eta)I_{\text{A,inj}}(t_1^{\eta_1})\prod_{j=2}^{2n} \eta_j H_{\text{A,inj}}(t_j^{\eta_j} ) \}\rangle - \int dt' \expval{I_{\text{A,inj}}(t')}\expval{I_\textrm{T}(0)}_{n-1} \\
= &  \frac{c_n}{4} \frac{(2n-1)!}{n!(n-1)!}(-i e^*)\sum_{\eta_0,\eta,\eta_j = \pm} \eta \prod_{j=2}^{2n}\eta_j\int dt dt_1 \cdots dt_{2n}    \Big[\langle T_K\{ I_\textrm{T}(0^{\eta_0})  H_\textrm{T}(t^\eta)\mathcal{A}(t_1^{\eta_1})\mathcal{A}^\dagger(t_2^{\eta_2})\prod_{i=2}^{n}\mathcal{A}^\dagger(t^{\eta_{2i-1}}_{2i-1})\mathcal{A}(t^{\eta_{2i}}_{2i})\} \rangle \\
& -\langle T_K \{ I_\textrm{T}(0^{\eta_0})H_\textrm{T}(t^\eta)\mathcal{A}^\dagger(t_1^{\eta_1})\mathcal{A}(t_2^{\eta_2})\prod_{i=2}^{n}\mathcal{A}^\dagger(t^{\eta_{2i-1}}_{2i-1})\mathcal{A}(t^{\eta_{2i}}_{2i}) \}  \rangle  \Big]  - \int dt' \expval{I_{\text{A,inj}}(t')}\expval{I_\textrm{T}(0)}_{n-1} \\
= & \frac{1}{4}\frac{e^*}{(n-1)!} \sum_{\eta_0,\eta = \pm}\eta\int dt\Bigg[(\Re[M-1]\frac{I_{\text{A,inj}}}{e^*}\abs{t} + i \Im[M-1]\frac{I_{\text{A,inj}}}{e^*}t  )^{n-1}\Big[\sum_{\eta_1,\eta_2=\pm}\eta_2\int dt_1dt_2\\
& \times  \big[-\expval{T_K\{\mathcal{T}^\dagger(0^{\eta_0})\mathcal{T}(t^\eta)\mathcal{A}(t_1^{\eta_1})\mathcal{A}^\dagger(t_2^{\eta_2})\}}+\expval{T_K\{\mathcal{T}^\dagger(0^{\eta_0})\mathcal{T}(t^\eta)\mathcal{A}^\dagger(t_1^{\eta_1})\mathcal{A}(t_2^{\eta_2})\}} \big] - 2\frac{I_\text{A,inj}}{e^*}\int dt'\expval{T_K\{ \mathcal{T}^\dagger(0^{\eta_0}) \mathcal{T}(t^\eta)\}}\Big] \\
&- (\Re[M-1]\frac{I_{\text{A,inj}}}{e^*}\abs{t} - i \Im[M-1]\frac{I_{\text{A,inj}}}{e^*}t  )^{n-1} \Big[\sum_{\eta_1,\eta_2=\pm}\eta_2\int dt_1dt_2\\
& \times  \big[-\expval{T_K\{\mathcal{T}(0^{\eta_0})\mathcal{T}^\dagger(t^\eta)\mathcal{A}(t_1^{\eta_1})\mathcal{A}^\dagger(t_2^{\eta_2})\}}+\expval{T_K\{\mathcal{T}(0^{\eta_0})\mathcal{T}^\dagger(t^\eta)\mathcal{A}^\dagger(t_1^{\eta_1})\mathcal{A}(t_2^{\eta_2})\}} \big] 
- 2\frac{I_\text{A,inj}}{e^*}\int dt'\expval{T_K\{ \mathcal{T}(0^{\eta_0}) \mathcal{T}^\dagger(t^\eta)\}}\Big]\Bigg] \\
=   &  \frac{e^* }{(n-1)!}\int_{-\infty}^\infty dt \expval{\comm{\mathcal{T}^\dagger(0)}{\mathcal{T}(t)}}_\text{eq} (\Re[M-1]\frac{I_{\text{A,inj}}}{e^*}\abs{t} + i \Im[M-1]\frac{I_{\text{A,inj}}}{e^*}t  )^n.
\end{split}
\end{equation}
Here $c_n = (-i)^{2n}/(2n-1)!$,
$I_{\text{A,inj}} =-ie^*(\mathcal{A}-\mathcal{A}^\dagger)$, $H_{\text{A,inj}} = \mathcal{A}(t) + \mathcal{A}^\dagger(t)$, $I_\textrm{T} = -ie^*(\mathcal{T} -\mathcal{T}^\dagger)$, $H_\textrm{T} = \mathcal{T} + \mathcal{T}^\dagger$.
In the third equality, the approximation valid at large $V_\text{inj}$ is applied as in Sec.~\ref{sec-dercorr} so that the $n!$ equivalent ways of pairing $t_{2i-1} \simeq t_{2i}$ are considered. %
The following was used in computing the last disconnected term of $\int dt' \expval{I_{\text{A,inj}}(t')}\expval{I_\textrm{T}(0)}_{n-1}$,
\begin{equation}
\begin{split}
\expval{I_\textrm{T}(0)}_{n-1} = \frac{1}{2(n-1)!} \sum_{\eta,\eta_0}\eta\int dt \Bigg[ & (\Re[M-1]\frac{I_{\text{A,inj}}}{e^*}\abs{t} + i \Im[M-1]\frac{I_{\text{A,inj}}}{e^*}t  )^{n-1} \expval{T_K\{\mathcal{T}^\dagger(0^{\eta_0})\mathcal{T}(t^\eta)\}}  \\ & - (\Re[M-1]\frac{I_{\text{A,inj}}}{e^*}\abs{t} - i \Im[M-1]\frac{I_{\text{A,inj}}}{e^*}t  )^{n-1} \expval{T_K\{\mathcal{T}(0^{\eta_0})\mathcal{T}^\dagger(t^\eta)\}}\Bigg].
\end{split} \label{cross-derv-disconnected1}
 \end{equation}
The diverging integral $\int dt'$ in the last term cancels (i.e., regularizes) divergence in other terms.
After calculations, the fourth equality is expressed with the equlibrium correlator $\expval{\comm{\mathcal{T}^\dagger(0)}{\mathcal{T}(t)}}_\text{eq}$.
Finally, %
we find $\expval{\delta I_{\text{A,inj} } \delta I_\textrm{T}}_n = e^* I_{\text{A,inj}} \partial \expval{I_\textrm{T}}_n/\partial I_{\text{A,inj} }$.
Collecting all the terms of different $n$, we prove Eq.~\eqref{AppA-result}. %
We note that a similar result was derived for Laughlin Abelian anyons using a different method of a non-equilibrium bosonization in Ref.~\cite{Rosenow16-S}.   
 
In the case of non-Abelian anyons having $\textrm{Im} [M] = 0$, we directly compute $\expval{\delta I_{ \text{A,inj}} \delta I_\textrm{T}}_n$ and $\expval{\delta I_{\text{A,inj}} \delta I_\textrm{T}}$, following the way in Sec.~\ref{subsection_subleading},
and find that the leading non-vanishing order of $\expval{\delta I_{\text{A,inj}} \delta I_\textrm{T}}$ is  $O\Big(I_\text{A/B,inj}^{2\delta+h_a-1}V_\text{A/B,inj}^{-h_a}\Big)$.
Hence, in this case $\expval{\delta I_{\text{A,inj}} \delta I_\textrm{T}}$ is negligible in comparison with 
$\expval{\delta I_\text{T}^2}$, as their ratio is $O\Big((\frac{\hbar}{e^{*2}}\frac{I_\text{A/B,inj}}{V_\text{A/B,inj}})^{h_a}\Big)$ as in Eq.~\eqref{order-fusa}.

\section{Side effects by anyons of the opposite chirality}\label{sec-back}
\begin{figure}[t]
	\centering
	\includegraphics[width = 0.42 \textwidth]{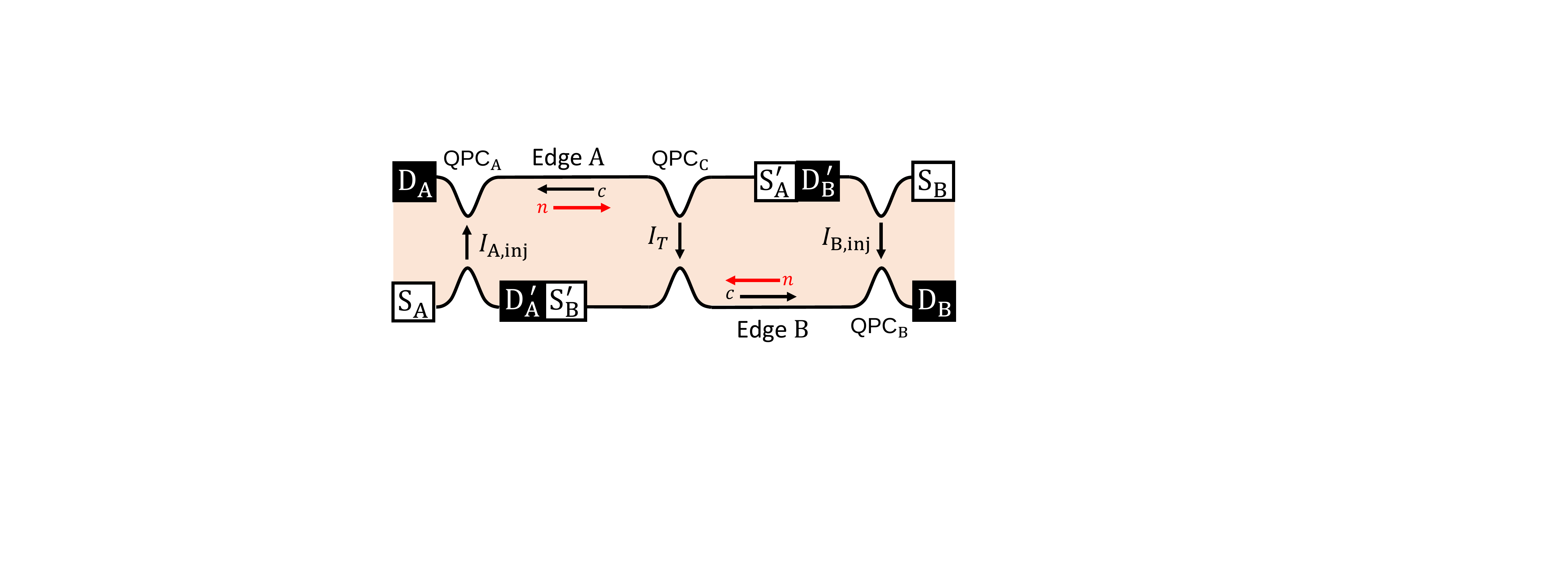}
	\caption{Anyon collider setup having for counter-propagating edge channels, upstream charge modes (black arrows, label $c$) and downstream neutral modes (red arrows, label $n$). The locations of the sources and detectors are different from Fig.~\ref{fig1}(a).
}\label{figs2-counter}
\end{figure}

At certain filling factors,
there appear downstream charge modes and upstream neutral modes together on quantum Hall edges.  In quantum Hall systems such as the anti-Pfaffian, particle-hole Pfaffian, and anti-Read-Rezayi states, it is expected that low-energy excitations of the upstream neutral modes are described by non-Abelian anyons.  Here we show that our result is directly applicable to the situation where downstream modes and upstream modes coexist on edges, when the collider QPC$_\textrm{C}$ is in the weak anyon tunneling regime.

Figure~\ref{figs2-counter} shows a collider for studying upstream neutral-mode anyons. This setup corresponds to that of Fig.~\ref{fig1}(a) in the main text, but having different locations of the sources and drains so that upstream neutral modes propagate from QPC$_\textrm{A/B}$ to QPC$_\textrm{C}$ on Edge A/B; in Fig.~\ref{fig1}(a) downstream charge modes flow from QPC$_\textrm{A/B}$ to QPC$_\textrm{C}$ in a reversed magnetic field.
It is possible to obtain information of the neutral modes from measurements of the charge currents and noises, since tunneling of neutral modes at QPCs is always accompanied by tunneling of charge modes.
 
When upstream and downstream modes coexist, a ``back-action" happens. In Fig.~\ref{figs2-counter}, neutral-mode anyons are injected to Edge A/B at QPC$_\textrm{A/B}$ and propagate to QPC$_\textrm{C}$. When tunneling of a neutral-mode anyon happens at QPC$_\textrm{C}$, it is accompanied by tunneling of a charge-mode anyon.  This charge-mode anyon propagates backward from  QPC$_\textrm{C}$ to  QPC$_\text{A/B}$, and it affects the injection current  $I_\text{A/B, \text{inj}}$ at QPC$_\textrm{A/B}$. In our parameter regime of $e^*V_{\alpha,\text{inj}}/\hbar  \gg I_{\alpha, \text{inj}}/e^* \gg I_\textrm{T}/e^*$, this back-action effect is negligible in comparison with $I_\text{T} $ and $\expval{\delta I_\text{T}^2}$. 
It is because the back-action results in an effective additional bias voltage of order of $2\pi \hbar I_\textrm{T}/e^{*2}$ across QPC$_\text{A/B}$ which is much smaller than $V_{\alpha,\text{inj}}$.

For concreteness, we compute the injection current $I_{\alpha, \text{inj} }$ at QPC$_{\alpha = \textrm{A,B}}$ in the presence of QPC$_\textrm{C}$. The back-action appears in the perturbation order $|\gamma_\textrm{C}|^2$ of tunneling at QPC$_\textrm{C}$ (while all the tunneling orders at QPC$_\alpha$ are considered),
\begin{equation} \begin{split}
I_{\alpha,\text{inj}}= \sum_{n=0}^\infty  \frac{(-i)^{2n+3}}{4(2n+1)!} 
  \sum_{\eta_0,\eta,\eta_a,\eta_b = \pm} \eta  \eta_a \eta_b \int dt dt_a dt_b \prod_{\eta_j} \eta_j\int dt_{j} \langle T_K\{ I_{\alpha,\text{inj}}(0^{\eta_0})H_{\alpha,\text{inj}}(t^\eta) H_\textrm{T}(t^{\eta_a}_a)H_\textrm{T}(t^{\eta_b}_b)\prod_{j = 1}^{2n}H_{\alpha,\text{inj}}(t^{\eta_{j}}_j)\} \rangle.	 
\end{split} \end{equation}
The integral over $t_j$'s is done at large $V_{\alpha,\text{inj}}$ as in Sec.~\ref{sec-dercorr}.  Expanded the integrand around $t\simeq 0$, we find
\begin{equation}
\begin{split}
I_{\alpha,\text{inj}} \simeq \frac{ e^*\abs{\gamma_\alpha^2}}{d_\psi} \sum_{\eta = \pm} \eta\int dt \frac{e^{-ie^*V_\text{inj}t/\hbar}}{[\epsilon - i \eta t]^{2\delta}} (1\mp i\frac{2\pi  \delta_\text{ch}}{e^*}I_\textrm{T} t).
\end{split}
\end{equation}
$\delta_\text{ch}$ is the tunneling exponent of the downstream charge mode part.
The second term is the back-action effect. Its sign factor $-$ (resp. $+$) is for $\alpha = \text{A}$ (resp. B).
We finally obtain
\begin{equation}
\begin{split}
I_{\alpha,\text{inj}} \simeq \frac{2\pi e^*\abs{\gamma_\alpha}^2}{d_\psi } \Big[\frac{1}{\Gamma(2\delta )}(\frac{e^*V_\text{inj}}{\hbar} )^{2\delta - 1}	 \pm \frac{1}{\Gamma(2\delta -1)}(\frac{e^*V_\text{inj}}{\hbar} )^{2\delta - 2}\frac{2\pi  \delta_\text{ch}}{e^*}I_\textrm{T}\Big].
\end{split}
\end{equation}
The correction term is of the order of $I_{\alpha,\text{inj}} \times \frac{I_\textrm{T} /e^*}{e^*V_\text{inj}/\hbar}$.
It is hence negligible in the regime of $(e^*)^2 V_{\alpha,\text{inj}}/\hbar  \gg I_{\alpha, \text{inj}} \gg I_\textrm{T}$.

\end{widetext}

\end{document}